%% file: main.tex
\documentclass[aps,floatfix,superscriptaddress,tightenlines,twocolumn,amsmath,amssymb,nofootinbib,raggedbottom,nobalancelastpage,notitlepage]{revtex4-1}

\input{formatting.tex}

\begin{document}

\title{\normalfont\Large{Optimised Domain-engineered Crystals for Pure Telecom Photon Sources}}

\author{A. Pickston}
\affiliation{Institute of Photonics and Quantum Sciences, Heriot-Watt University Edinburgh, UK, EH14 4AS}

\author{F. Graffitti}
\affiliation{Institute of Photonics and Quantum Sciences, Heriot-Watt University Edinburgh, UK, EH14 4AS}

\author{P. Barrow}
\affiliation{Institute of Photonics and Quantum Sciences, Heriot-Watt University Edinburgh, UK, EH14 4AS}

\author{C. Morrison}
\affiliation{Institute of Photonics and Quantum Sciences, Heriot-Watt University Edinburgh, UK, EH14 4AS}

\author{J. Ho}
\affiliation{Institute of Photonics and Quantum Sciences, Heriot-Watt University Edinburgh, UK, EH14 4AS}

\author{A. M. Brańczyk}
\affiliation{Perimeter Institute for Theoretical Physics, Waterloo, Ontario, Canada, N2L 2Y5}

\author{A. Fedrizzi}
\affiliation{Institute of Photonics and Quantum Sciences, Heriot-Watt University Edinburgh, UK, EH14 4AS}

%%%%%%%%%%%%%%%%
%%% abstract %%% 
%%%%%%%%%%%%%%%%

\begin{abstract}
\vspace{1em}
The ideal photon-pair source for building up multi-qubit states needs to produce indistinguishable photons with high efficiency. 
Indistinguishability is crucial for minimising errors in two-photon interference, central to building larger states, while high heralding rates will be needed to overcome unfavourable loss scaling. 
Domain engineering in parametric down-conversion sources negates the need for lossy spectral filtering allowing one to satisfy these conditions inherently within the source design.
Here, we present a telecom-wavelength parametric down-conversion photon source that operates on the achievable limit of domain engineering. 
We generate photons from independent sources which achieve two-photon interference visibilities of up to $98.6\pm1.1\%$ without narrow-band filtering.
As a consequence, we reach net heralding efficiencies of up to 67.5\%, which corresponds to collection efficiencies exceeding $90\%$.
\end{abstract}
\maketitle

%%%%%%%%%%%%%%%%%%%%
%%% introduction %%% 
%%%%%%%%%%%%%%%%%%%%

Scalable photonic quantum technologies require pure photons created on demand. 
The simplicity of using photon sources based on spontaneous parametric down-conversion (PDC) means the process has been exploited widely and studied in great depth \cite{CHRIST2013351,slussarenko2019photonic}. 
Efforts have been made to achieve pseudo-deterministic operation via multiplexing~\cite{PhysRevA.66.042303, PhysRevA.66.053805, Kanedaeaaw8586, Collins2013, Kiyohara16muxWithPNR, Francis-Jones16muxFibre, Ma2011muxEOMs, Mendoza16mux, Broome:11, doi:10.1063/5.0003320}, reach high heralding efficiencies, and generate indistinguishable photons---characteristics that all contribute towards an ideal source of photons. 
Whilst deterministic operation can be addressed separately, photon source engineering must focus on generating indistinguishable photons with high heralding efficiencies, since tasks such as measurement-based quantum computing \cite{PhysRevA.68.022312,oneWay}, photonic Boson sampling \cite{Broome794,PhysRevA.101.063821} and photonic quantum repeaters \cite{allOptical} are ultimately contingent on high visibility two-photon interference at high rates with minimal losses.

Our work focuses on tailoring the phase matching function (PMF), modifying the PDC interaction to produce optimal photons. 
The quantum state resulting from PDC, when considering solely terms which describe emission of a single pair reads,
\begin{equation}\label{eq:simplePair}
    \ket{\psi}_{\text{pair}} =  \iint d\omega_s d\omega_i  f(\omega_s,\omega_i) \hat{a}_s^{\dagger}(\omega_s) \hat{a_i}^{\dagger}(\omega_i) \ket{0}.
\end{equation}
The state contains $f(\omega_i,\omega_s)$, which is determined by the pump envelope function (PEF) $\alpha(\omega_s + \omega_i)$, and PMF $\phi(\omega_s,\omega_i)$,
\begin{equation}
    f(\omega_s,\omega_i) =  \phi(\omega_s,\omega_i) \hspace{1pt} \alpha(\omega_s + \omega_i),
\end{equation} 
and is referred to as the Joint Spectral Amplitude (JSA). Under symmetric group-velocity matching---where the mean of the inverse signal-idler group velocities are matched to the inverse of the pump group velocity~\cite{PhysRevA.64.063815,u2006generation,PhysRevLett.100.133601,Jin:14,Jin:13,Greganti:18}---the PMF and PEF are orthogonal.
In this condition, signal and idler photons can be interfered interchangeably, unlike in Ref. \cite{PhysRevLett.117.210502}, as well as in heralded single-photon schemes.

\begin{figure*}[t]
    \vspace{0em}
    \begin{center}
    \includegraphics[width=1.5\columnwidth]{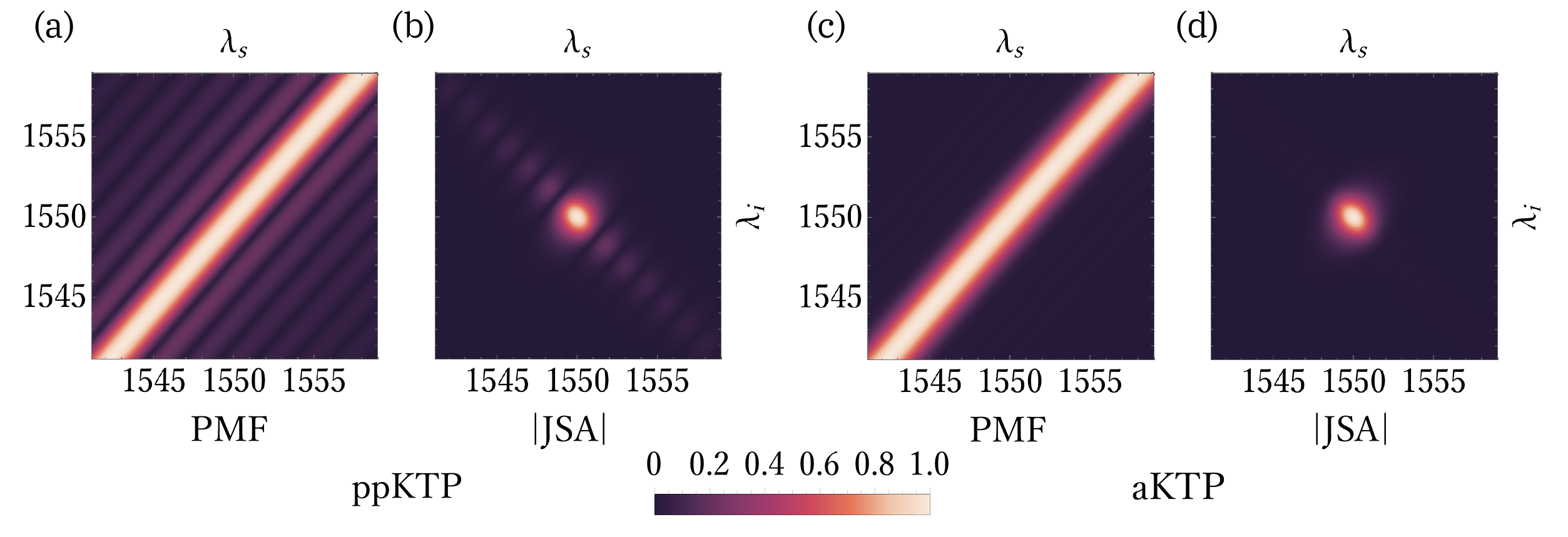}
    \vspace{-1em}
    \end{center}
    \caption{\textsf{}Theoretical Phase Matching Functions and Joint Spectral Amplitudes for periodically poled (a, b) and our Gaussian apodized crystals (c, d). The prevalent correlations in the joint spectrum for the periodically poled crystal are a result of the Sinc shaped phase matching function (a) and must be filtered out with narrow-band filters to achieve high spectral purity. These correlations are suppressed in the apodized crystals joint spectrum (compare (b) and (d)) by targeting a Gaussian phase matching function (c), increasing spectral purity, increasing source indistinguishability and removing the need for tight spectral filtering.}
    \label{fig:expJSIandMarginal}
\end{figure*}

Achieving unit photon purities requires the bi-photon states to exist in a single spectral mode.
In standard non-linear crystals, the PMF is a sinc-shaped function, which generates spectral correlations in the JSA, leading to bi-photon states that exist in a superposition of spectral modes~\cite{Branczyk:11,graffitti2018design}, illustrated in Figure~\ref{fig:expJSIandMarginal}~(a, b).
These correlations  reduce spectral photon purity and thus indistinguishability.
Typically, tight filtering is used to suppress the spectral correlations, increasing purity and interference visibility.
But filtering introduces optical loss, leading to a reduction in heralding efficiencies, source brightness and  photon-number purity \cite{Bra_czyk_2010,PhysRevA.95.061803}.
One can achieve a factorable JSA without tight filtering however, by engineering the properties of the crystal such that the PMF approximates a Gaussian function, shown in Figure~\ref{fig:expJSIandMarginal}~(c).
First suggested by Brańczyk et al.~in Ref.~\cite{Branczyk:11}, several methods for obtaining a Gaussian PMF have been developed.
Altering the poling duty cycle of the crystal domains~\cite{BenDixon:13,Chen:17,Chen:19,PhysRevApplied.12.034059}, the orientation of the poling direction~\cite{Dosseva}, and tailoring both~\cite{graffitti2018independent,Tambasco:16,graffitti2017pure} can all generate the desired function. 
Using an optimal technique developed in Ref.~\cite{graffitti2017pure}, Graffitti et al.~demonstrated interference of photons generated from independent domain-engineered crystals in Ref.~\cite{graffitti2018independent}. 
Within that work, a symmetric heralding efficiency of 65\% was achieved along with a source brightness of 4kHz/mW and lower-bound photon purity of 90.7$\pm$0.3\%.
While developed primarily for generating separable photons, domain engineering can also be exploited for tailoring high quality non-Gaussian PMFs, e.g. for efficient generation of time-frequency mode entanglement~\cite{PhysRevLett.124.053603} and time-frequency hyper-entanglement~\cite{graffitti2020hyperentanglement}.

Here we present a PDC source based on domain-engineered crystals, operating on the achievable limits of this technique.
Through the optimisation of parameters which trade off the non-trivial relationship between non-linearity and indistinguishability, we establish a lower bound on spectral purity of $98.0\pm0.1\%$, achieve a maximal visibility of $98.6\pm1.1\%$, a symmetric heralding efficiency of $67.5$\% and a source brightness of $4.1$~kHz/mW.

%%%%%%%%%%%%%%
%%% Theory %%% 
%%%%%%%%%%%%%%

PDC occurs in non-centrosymmetric optical materials, such as potassium titanyl phosphate (KTP).
Quasi-phase-matching (QPM), a method commonly used to bypass non-critical phase-matching, is achieved by inverting the orientation of the crystal lattice structure with a period that prevents destructive interference of signal and idler fields.
This allows photon generation along the crystallographic axes, thus avoiding birefringent walk-off effects and permits photon generation at desired wavelengths \cite{161322}.
The non-linear response of a uniformly periodically-poled crystal corresponds to a step function, which, in the Fourier domain, transforms to the detrimental sinc function seen in Figure~\ref{fig:expJSIandMarginal}~(a).

\begin{figure*}[t]
    \begin{center}
    \includegraphics[width=1.5\columnwidth]{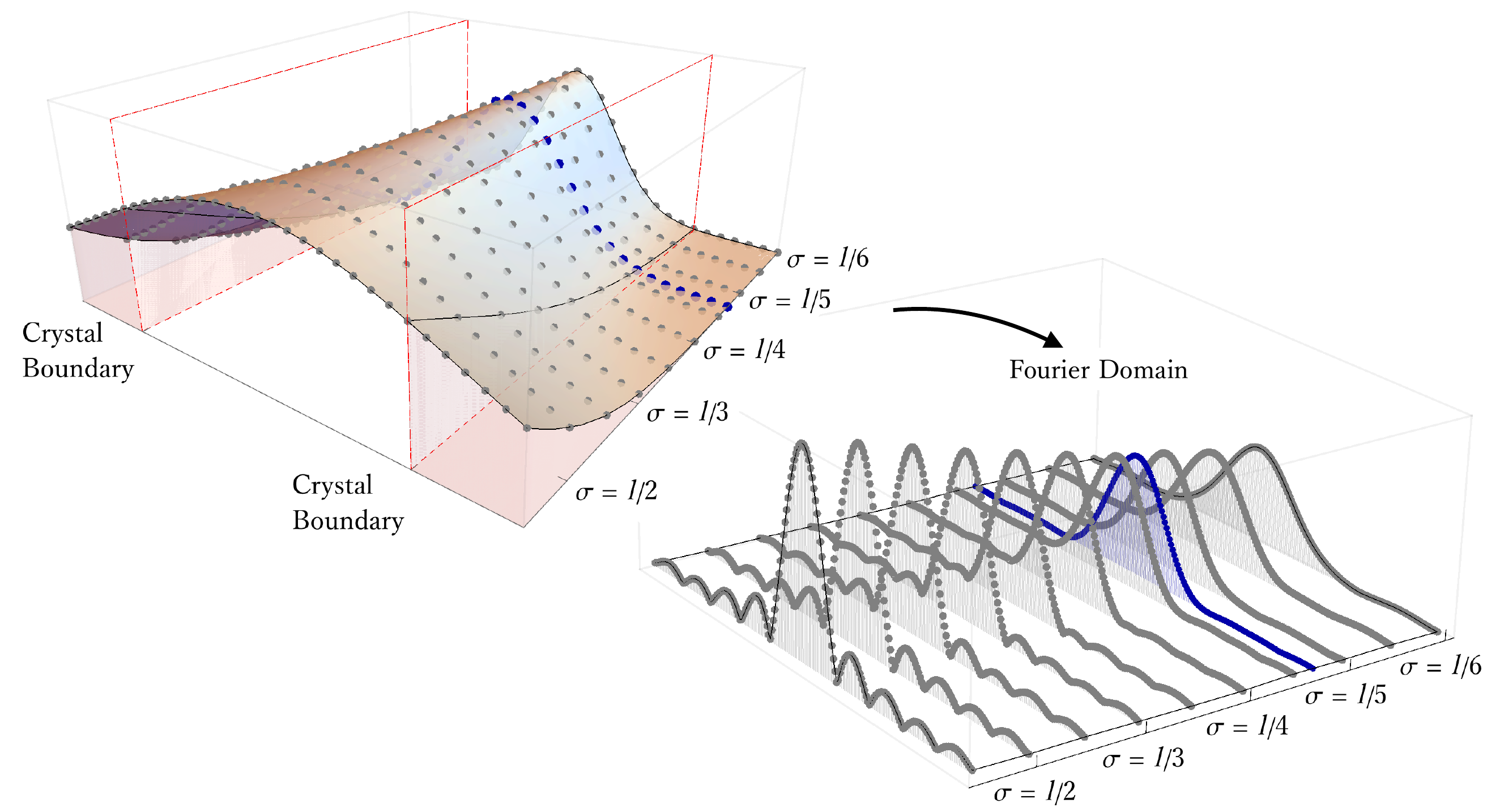}
    \vspace{-1em}
    \end{center}
    \caption{\textsf{}Target function for Gaussian domain engineering. The top panel shows the target function of varying widths, with the red shaded areas indicating regions outside the boundary fixed by the crystal length $l$. A target function that is too wide, for example when $\sigma=l/2$, will result in side lobes in the PMF which is shown on the bottom panel. A narrow target function may produce minimal side lobes in the PMF, but will result in a lower effective non-linearity and therefore a lower source brightness. The blue dotted lines indicate the trade-off we chose for our implementation.}
    \label{fig:sigmaPMF}
\end{figure*}

To achieve high purities in the group-velocity matching (GVM) regime, the pump envelope function should be a transform-limited Gaussian envelope, whilst the phase-matching function should also be a Gaussian function \cite{graffitti2018design,PhysRevA.98.043813}.
Typical mode-locked lasers have a sech$^2$-shaped PEF, which in our case contributes $\sim$1\% in purity decrease. In this work we focus on PMF engineering only, but it's also possible to reshape the pump field spectrum into a Gaussian PEF, as recently demonstrated by C.~Chen \textit{et al.}  \cite{Chen:19}.

We define the PMF as:
\begin{equation}\label{eq:constantsTheta}
    \phi (\omega_s,\omega_i) = \int^{+\infty}_{-\infty} \hspace{1pt} \text{g} (z) \text{e}^{i \Delta k(\omega_s,\omega_i)z} \text{d}z,
\end{equation}
where $\Delta k(\omega_s,\omega_i) = k_p (\omega_s + \omega_i) - k_s(\omega_s) - k_i(\omega_i)$ is the phase mismatch arising from material dispersion and $\text{g} (z) = \chi^{(2)}(z)/\chi^{(2)}_0$ is the normalised non-linear response of the material.
We can modify this function by aperiodically altering the width and orientation of each domain while tracking a predefined target function $\text{g} (z)_{\text{target}}$~\cite{graffitti2017pure}. 
This target function produces a target PMF amplitude which is scaled to possess the maximum gradient achievable---that is $\frac{\pi}{2}$~\cite{boyd}---to ensure that the non-linear response along the longitudinal direction is maximised. 
The resulting PMF amplitude \cite{graffitti2017pure,Tambasco:16} is given by:

\begin{equation}    
    \text{PMF}(\Delta k_0) = \sqrt{\frac{\pi }{2}} \left(\text{erf}\left(\frac{l}{2 \sqrt{2} \sigma }\right)+\text{erf}\left(\frac{z-\frac{l}{2}}{\sqrt{2} \sigma }\right)\right).
\end{equation}

A crucial parameter is the choice of $\sigma$, the width of the Gaussian function.
This parameter balances source brightness with spectral purity.
In order to obtain high brightness the function must be wide, but to avoid correlations the function must be narrow.
Thus we choose a width that both avoids a large step in non-linearity---avoiding spectral correlations---whilst wide enough to obtain a reasonably high effective nonlinearity and thus brightness.
This trade-off is illustrated in Figure~\ref{fig:sigmaPMF}.
With a $\sigma =  l/4.7$, where $l$ is the crystal length, the generation of side lobes is minimal whilst not adversely reducing generation rates, see Figure~\ref{fig:expJSIandMarginal}~(c) for our apodized crystals theoretical PMF. 
The crystal length is $l = 30$ mm, resulting in $\sigma = 6.38$ mm.

%%%%%%%%%%%%%%%%%%
%%% Experiment %%% 
%%%%%%%%%%%%%%%%%%

\begin{figure*}[t]
    \centering\includegraphics[width=12.5cm]{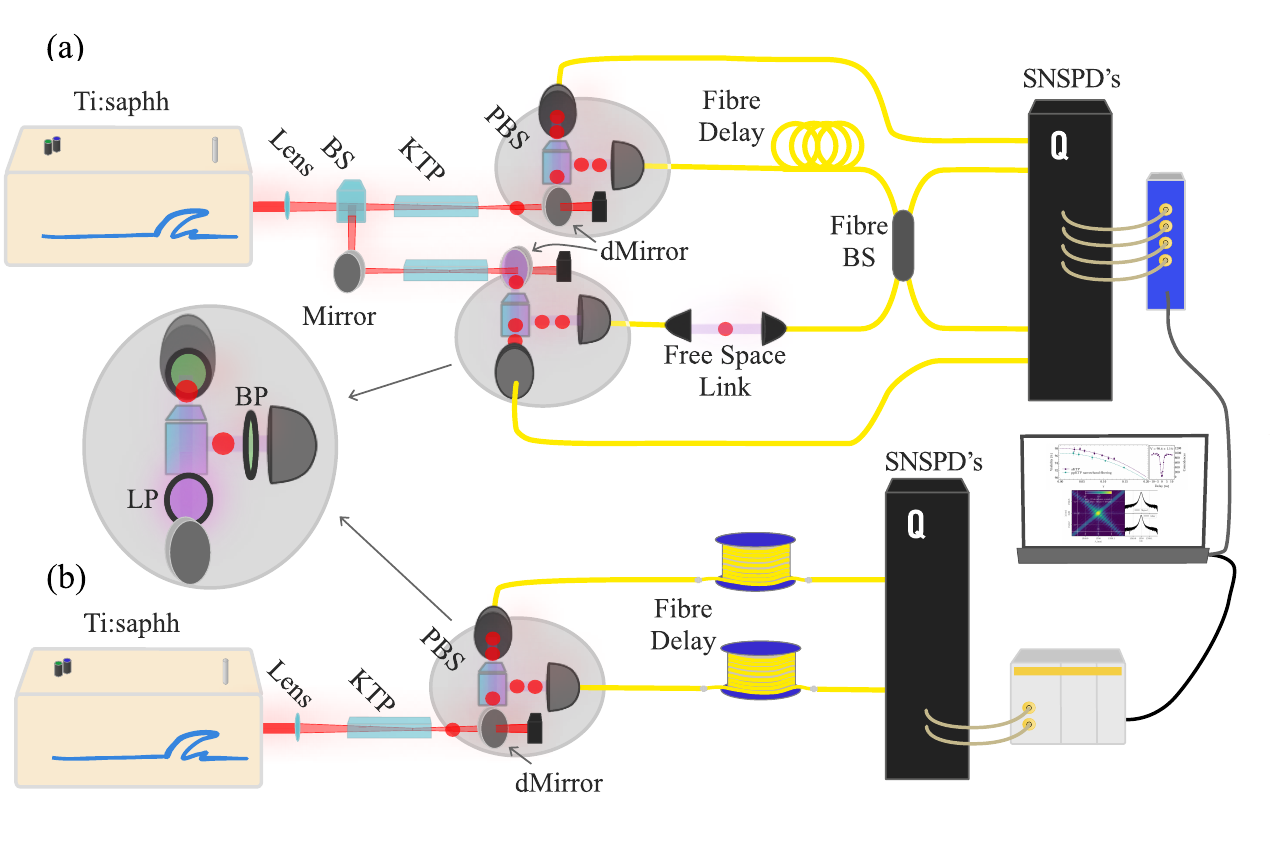}
    \caption{\textsf{}Experimental Layout. (a) A Ti:Saphh laser pumps a standard ppKTP, or domain-engineered aKTP crystal, at a repetition rate of 80.9MHz. The down-converted photon pairs are collected after reflection from a dichroic mirror and separated by a PBS. Individual photons from two sources are temporally synchronised with an adjustable free-space gap before they are superposed in an in-fibre BS. Photons are then detected by Superconducting Nano-wire Single Photon Detectors (SNSPDs), with photon arrival times being time-tagged and processed. \textsf{}(b) Two $\sim20$km fibre spools of telecommunication grade fibre are used for dispersive spectroscopy, exploiting chromatic dispersion allowing us to reconstruct the joint photon spectrum \cite{Avenhaus:09}. We collect the photon pairs in the same manner as above, however the collected photons are subjected to the fibre delay.}
    \label{fig:layout}
\end{figure*}

Using a mode-locked Ti:Sapphire laser with a non-ideal $\text{sech}^{2}$-shaped spectral envelope we pump our crystals at a wavelength of 774.9~nm, down-converting to 1549.8~nm. 
The pulse duration can be tuned between 1.3~ps and 1.4~ps to be matched to different crystal and filter combinations. 
Operating at just below $1550$~nm was necessary to ensure temperature stabilisation, enabled by keeping our crystals sufficiently far from room temperature for degenerate photon generation. 
We focus into the centre of the crystal with a 40~cm lens, generating a slightly elliptical spot with a waist of $\sim124\mu$m in the horizontal and $\sim119\mu$m in the vertical axis. 
This focusing condition was chosen as an optimal trade-off between brightness and heralding efficiency~\cite{BenninkRyanS2010OcGb,PhysRevA.83.023810,PhysRevA.90.043804}. 
To collect the down-converted modes we separate the emitted photon pairs on a polarising beam splitter, with an initial dichroic mirror removing pump photons. 
Signal and idler photons are collected into single-mode fibres after long-pass filtering to reduce any residual pump photons further.
We introduce some gentle filtering around the central spectral lobe of our down-converted photons via a filter with a transmission profile of $\text{exp}[-\frac{(\omega - \omega_0)^4}{2\sigma^4}]$, a FWHM of 7.4~nm and is $\sim$5 times wider than the generated photon bandwidth which minimally impacts heralding efficiencies. 
Down-converted photons then pass through optical interference or spectroscopy setups before being collected by Superconducting Nano-wire Single Photon Detectors (SNSPDs) operating at $\sim80\%$ detection efficiencies.
See Figure~\ref{fig:layout}~(a) for the experimental layout.

\begin{figure*}[t]
    \vspace{-1em}
    \centering\includegraphics[width=11.5cm]{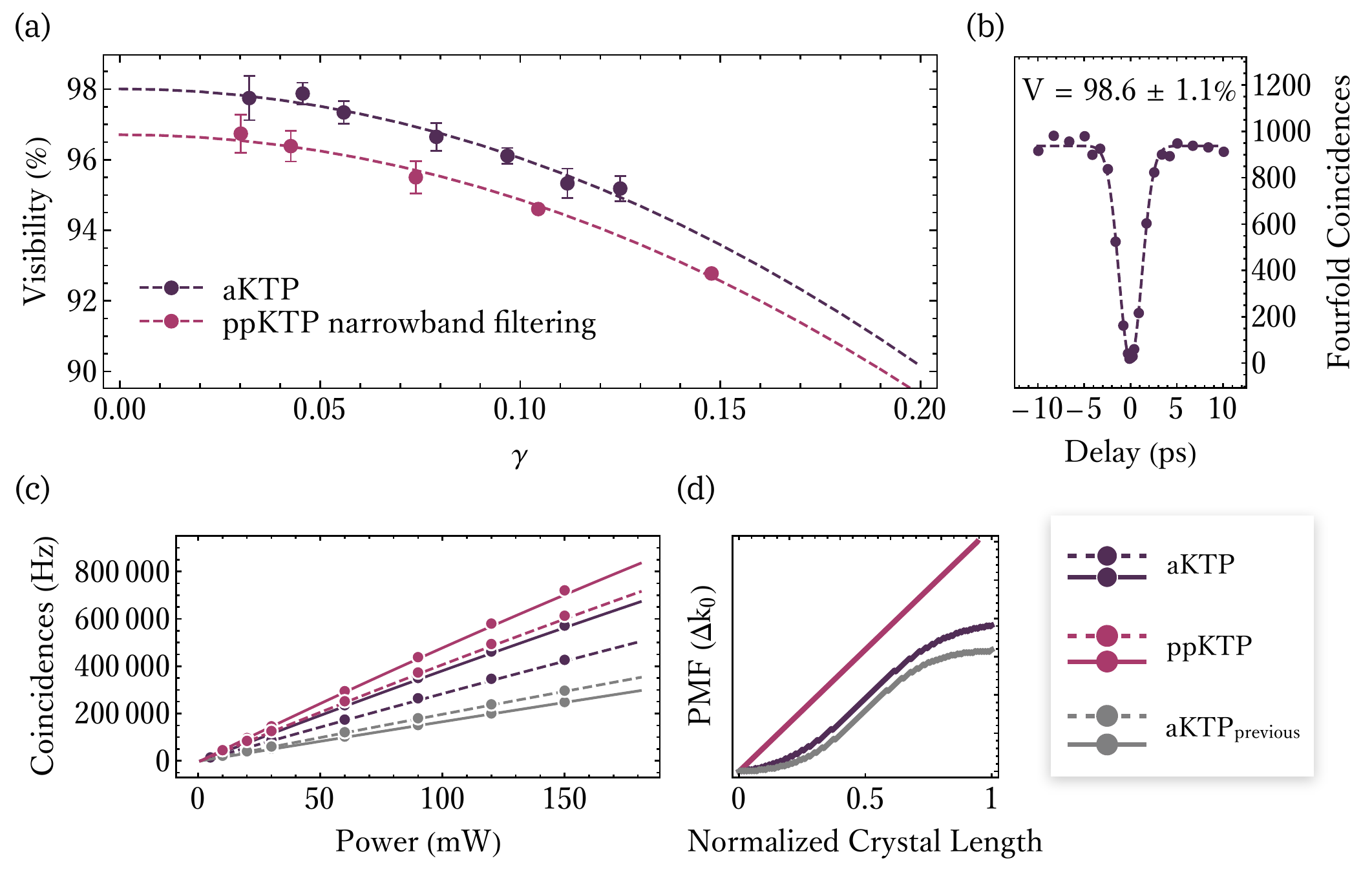}
    \caption{Experimental results from interfering photons generated from two independent sources. (a) Visibility dependence on the squeezing parameter, $\gamma$. Each data point represents the average visibility from five interference measurements for each value of pump power (or, equivalently, for each value of $\gamma$). From this data set we can infer a minimum spectral purity of $98.0\pm0.1\%$ and compare the performance of our aKTP crystals with loose spectral filtering against a ppKTP crystal with narrow-band, tight spectral filtering. (b) A two-photon interference measurement between photons generated from separate sources. At a pump power of 10~mW we achieve an interference visibility of $98.6\pm1.1\%$, with a four-photon coincidence rate of around 5~Hz. (c) Photon pair generation rates of our previous crystal, a filtered ppKTP crystal and our aKTP crystal. (d) Theoretical amplitude of the phase matching function along the longitudinal direction (z axis) of the crystal at $\Delta k_0$.}
    \label{fig:visibility}
\end{figure*}

We investigated two-photon interference visibilities for different configurations of crystals---a 22~mm periodically-poled KTP crystal and a 30~mm custom-poled KTP crystal---and filters.
We interfered photons generated from separate, but identical (manufactured from the same lithographic mask) crystals.
In order to obtain a lower bound on the implied photon purity and to generate the data in Figure~\ref{fig:visibility}~(a), the two sources were pumped with the same amount of pump power and at least five independent two-photon interference scans were run consecutively. 
The data acquisition time for each of these scans was sufficient to obtain at least 1000 four-photon coincidence events outside of the dip position. 
From this data set we fit a linear function and extrapolate the expected visibility at zero pump power. This technique eliminates visibility degradation due to photon-number impurity (see the Appendix of Ref.~\cite{graffitti2018independent}) and serves to lower bound photon purity. 
The performance of all results are summarised in Table~\ref{tab:data}. 
The different crystal photon generation rates, in terms of number of coincident photon counts per second, are shown in Figure~\ref{fig:visibility}~(c). 
Importantly, the generation rates and heralding efficiencies are quoted with consistent focusing conditions in the same optical setup and provide a comparison and not an upper limit on crystal performances. 
Different pump focusing conditions as well as different collection optics will result in different values for source brightness, heralding efficiencies and can also impact photon purity~\cite{BenninkRyanS2010OcGb}. 

\begin{table*}[t]
    \small
    \centering\setlength{\extrarowheight}{-2pt}
    \begin{tabular}{r@{\hskip 0.2in} c@{\hskip 0.2in} c@{\hskip 0.2in} c@{\hskip 0.2in} c@{\hskip 0.2in} c@{\hskip 0.2in} c@{\hskip 0.2in}} 
    \toprule \\[-1ex]
    
    Crystal     &        \begin{tabular}[c]{@{}c@{}c@{}}Interference \\Visibility\\ (\%)\end{tabular}      &     \begin{tabular}[c]{@{}c@{}c@{}}Mean \\ Heralding \\Efficiency \\(\%)\end{tabular}      &       \begin{tabular}[c]{@{}c@{}c@{}}Collection \\Efficiency \\ (\%)\end{tabular}     &       \begin{tabular}[c]{@{}c@{}c@{}}Mean \\ Brightness \\ (cc/mW/s)\end{tabular}     &       \begin{tabular}[c]{@{}c@{}c@{}} Experimental\\$\sqrt{\text{JSI}}$\\Purity\\ (\%)\end{tabular}       &       \begin{tabular}[c]{@{}c@{}c@{}}Theoretical\\JSA\\Purity\\ (\%)\end{tabular}\\[-1ex]  \\
    
    \hline\hline\\
    
    aKTP & 98.0 $\pm$ 0.1 & 67.5 & $91.8$ & 3900 & 91.17 $\pm$ 0.02 & 98.7 \\[1ex]
    
    ppKTP & 95.3 $\pm$ 0.1 & 57.2 & $77.4$ & 4900 & 94.43 $\pm$ 0.03 & 98.4 \\[1ex]
    
    \bottomrule
    
    \end{tabular}
    \caption{A summary of results comparing our custom aKTP crystal with loose spectral filters to a ppKTP crystal with tight spectral filters. The interference visibilities are quoted at zero pump power. The mean heralding efficiencies and brightness respectively for each crystal result from an analysis of the performance of each source as a function of pump power. The collection efficiencies are calculated with respect to the upper limit detection efficiency of our detectors (80\%) as well as other known optical losses (7.9\%). Finally we also include the purities calculated from our experimental JSI reconstructions, as well as the theoretical purities. We use the $\sqrt{\text{JSI}}$ to calculate purities as it represents a better approximation of the JSA compared to calculating the purity of the JSI \cite{graffitti2018design}.} 
    \label{tab:data}
    \vspace{-1em}
\end{table*}

%%%%%%%%%%%%%%%
%%% Results %%% 
%%%%%%%%%%%%%%%

We observe an improvement in both interference visibility and generation rates upon Ref.~\cite{graffitti2018independent}, a result of altering the width of the Gaussian target function tracked by our algorithm from $\sigma = l/4$ to $\sigma = l/4.7$.
Ref.~\cite{graffitti2018independent} reported a lower bound purity of $92.7\pm0.2\%$. This data was obtained using a delayed two-photon interference technique, interfering photons generated from the same source. Instead of this technique, we perform interference measurements on photons from independent crystals, representing a true proxy for source scalability. 
Our new apodized crystals have a lower-bound purity, under the same gentle filtering, of $98.0\pm0.1\%$. Without any filtering we obtain a lower-bound purity of $95.3\pm 0.1\%$ and the respective data contributes to a full plot of all results found in the Appendix. 

Rather than expressing results in terms of pumping power, we show the main results in terms of $\gamma$, the effective squeezing of the two-mode squeezed vacuum, which encompasses the pump power and focusing conditions applied to the crystal. 
In the photon number basis, $n$, the PDC process can be expressed as $(1-\gamma^{2})^{1/2}\sum^{\infty}_{\text{n} = 0}\gamma^{\text{n}}\ket{\text{n},\text{n}}_{s,i}$, with $\gamma$ defined as: $\gamma=(\tau\text{p})^{1/2}$, where p is the pump power and $\tau$ is a constant quantifying the non-linear interaction of the medium~\cite{JIN201547}.
In this work, we evaluate \(\gamma\) from the measured coincidence rates, single rates and the clock rate of the pulsed laser in a similar manner as in Ref.~\cite{JIN201547}.
With knowledge of \(\gamma\), the photon pair rate and multi-photon pair rates can be determined.
This forms a more representative analysis of crystal performance as the variety of experimental conditions distinct to our analysis are gathered into this one term. 
Figure~\ref{fig:visibility}~(a) therefore, compares the interference visibility of our aKTP crystals performance with a ppKTP crystal as a function the squeezing, $\gamma$. 

With apodization, the need for tight filtering is removed, resulting in significantly higher heralding efficiencies, seen in Table \ref{tab:data}. 
This higher efficiency means that when both sources are generating photons at the same raw rate, the source with higher heralding efficiencies will lead to higher rates of detector clicks. 
Factoring out known optical losses and detection efficiencies (taken as the quoted operational upper bound of 80\%), overall collection efficiencies are lower bounded to 91.8\%. 
Optical losses were determined by measuring the transmission properties of each optical element between pair production and the housing of our detectors, this accounted for a loss of 7.9\%. 
Anti-reflection coated optics were used where possible to minimise any losses, including on the end facets of all the KTP crystals used in this investigation.

\begin{figure*}[t]
    \centering\includegraphics[width=1.4\columnwidth]{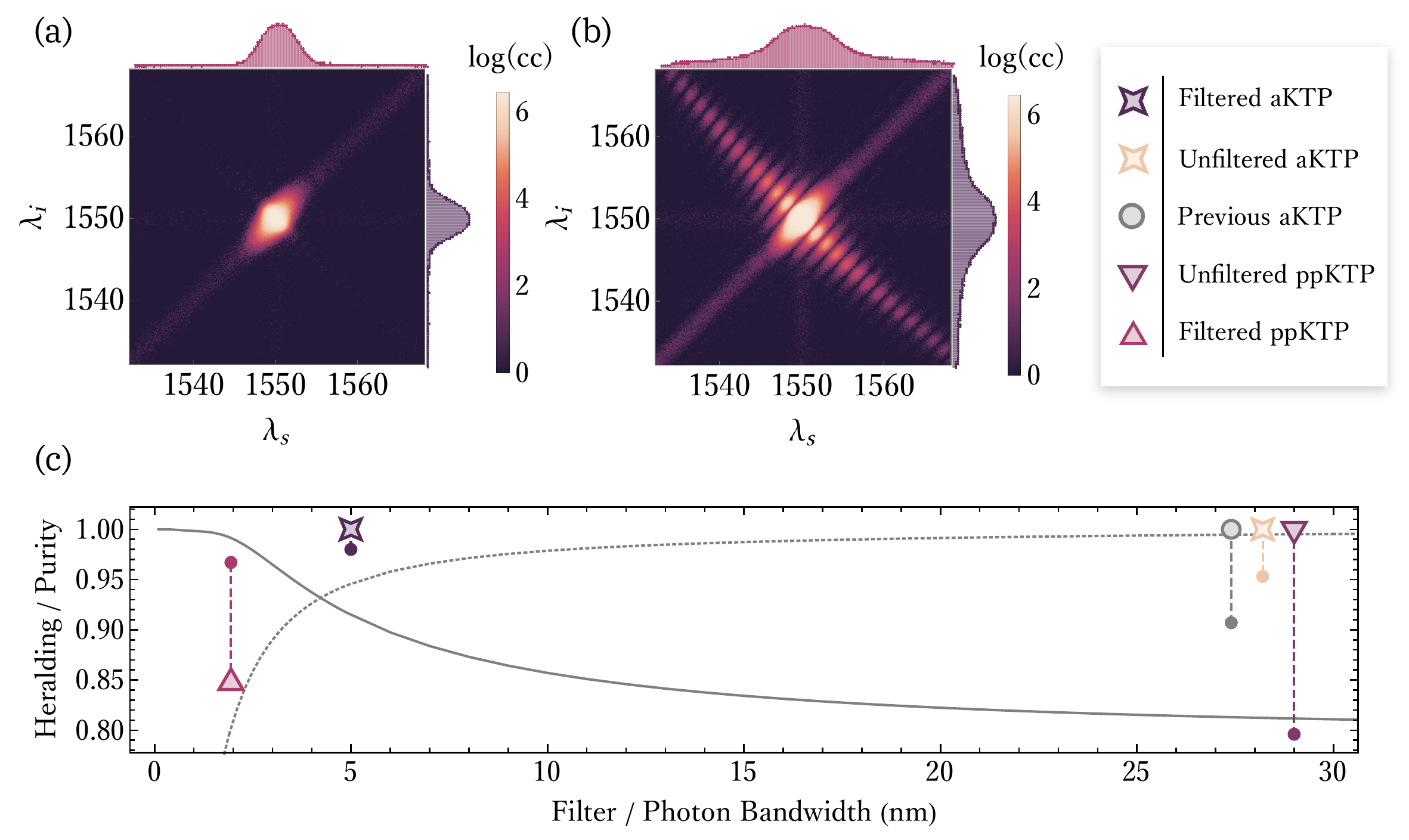}
    \caption{(a) and (b) Experimental reconstruction of the JSI and marginal photon spectrum. Using a dispersive spectroscopy technique, we construct the full joint spectrum spanning a whole repetition of our lasers cycle, for our aKTP crystal (a) and ppKTP crystal (b). The reconstruction reveals all spectral correlations which are then either suppressed by filtering, or already suppressed through modification of the PMF. These figures are plotted with a logarithmic scale in order to highlight any correlations. (c) Normalised heralding and purities of the crystals we have analysed in this manuscript as a function of the photon bandwidth or the filtered photons bandwidth. The solid data points represent the normalised heralding, the filled data points are purity values and the solid (dashed) lines are the simulated results of the heralding (purity) for the ppKTP crystal.}
    \label{fig:spectraExp}
\end{figure*}

Another means of quantifying source performance is to analyse a reconstruction of the joint photon spectrum. 
Reconstruction of the JSA is experimentally demanding since it requires a spectrally resolved amplitude and phase measurement, which can be achieved for example via phase-sensitive stimulated emission tomography~\cite{Jizan:16}. 
Constructing the joint spectral intensity (JSI), equivalent to $|{\text{JSA}}|^2$, can be achieved with comparative ease and is therefore commonly shown, although one has to be careful what conclusions to draw in the absence of phase information normally contained in the JSA~\cite{graffitti2018design}. 
With 20~km of standard telecommunication fibre optic we can exploit chromatic dispersion to map photon arrival time to the associated spectral component of the JSI, as performed in~\cite{PhysRevLett.124.053603}. 
The experimental arrangement is depicted in Figure~\ref{fig:layout}~(b). Collection of at least $10^6$ photons detected by SNSPD's operating with $<50$~ps jitter, $<25$~ns reset time and processed via a Picoquant HydraHarp with 1~ps timing resolution, enabled the construction of the respective JSI for combinations of filter and crystal. 
The spectral window of our results span 12.5~ns and the achievable timing resolution of this spectrometer translates to a spatial resolution of $\sim0$.0028~nm.

Figure~\ref{fig:spectraExp}~(a) and (b) shows the respective experimental JSIs of un-filtered aKTP and un-filtered ppKTP on a logarithmic scale. 
Any spectral correlations that exist along the main diagonal are visually highlighted. These correlation are clearly prevalent for ppKTP but almost non-existent for unfiltered aKTP, a result of non-zero contributions from the PMF. 
Along the diagonal, from bottom left to the top right, as well parallel to the x and y axes, through the central lobe of the joint spectra, we see a constant background signal arising from dark counts. 
An additional PDC source was used as a trigger, to measure the arrival of signal and idler photons. 
A dark count detected in the trigger channel, as opposed to an actual photon, corresponds to a displacement of the central lobe along the diagonal, resulting in temporally correlated background noise.
If, either the signal or idler photon is lost, but a dark count is detected in that channel along with the trigger and remaining signal/idler photon, the central lobe is shifted in the parallel to the x or y axes depending on which of signal and idler photons are lost. 
The probability of this is smaller, proportional to the pair emission probability.

To produce estimates for both the JSI and $\sqrt{\textrm{JSI}}$ purities, we reconstructed the JSI across increasingly long measurement intervals. 
Each estimation is calculated using $50\times50~\textrm{ps}$ bins; doing so reduces the sparsity of the raw data and provides a more accurate and reliable Singular Value Decomposition (SVD). 
The SVD is used to numerically implement the Schmidt decomposition, used to quantify the non-separability of the JSA~\cite{PhysRevLett.84.5304}. 
By observing the value the estimation converges towards, we truncate the total measurement time to avoid instability. These estimation of purities are contained in Table \ref{tab:data}. 
Neither the JSI nor the $\sqrt{\text{JSI}}$ truly reveal photon purity due to lack of phase information, something two-photon interference can incorporate~\cite{graffitti2018design}. 
Thus, two-photon interference results represent a more faithful estimate of photon purity. 
Discrepancies between the lower-bound purities determined by two-photon interference results, and inferred purities from experimental JSIs could be caused by a combination of different factors, such as drifts in the laser central frequency and pulse duration, as well as non-negligible jitter in the detection system. 
Visually noticeable elongation of central lobe along the diagonal suggests pump pulse durations that are shorter than the crystal is optimised for, which in turn would result in a lower purity for experimental JSI analysis. 
From simulations we estimate that, pulse durations that are $\pm0.4$~ps away from the ideal value can result in purities dropping by 6\%, see the Appendix for more details.

%%%%%%%%%%%%%%%%%%
%%% Discussion %%% 
%%%%%%%%%%%%%%%%%%

The importance of achieving the photon source characteristics displayed in this work was recently highlighted in Ref.~\cite{PhysRevA.101.063821}, which concludes that quantum supremacy in a Boson sampling task with non-multiplexed single-photons from PDC sources can only be achieved with Gaussian-profile aKTP crystals due to the higher collection efficiencies.
Notably, photonic quantum supremacy has just been demonstrated in Gaussian Boson Sampling (GBS), in an experiment which created 50 photons from 25 Gaussian apodized crystals using a duty-cycle poling technique~\cite{Zhong1460}. 
Using our improved poling algorithm and considering the trade off between non-linearity and photon purity highlighted in this manuscript, an optimal $\sigma$ could enable higher purities and heralding efficiencies.
This, in turn, would culminate in a greater advantage and scalability of the scheme.

The discrepancy in brightness between our aKTP source and the ppKTP source highlighted within Table \ref{tab:data} can be balanced by adjusting the relative pump powers to achieve the same squeezing $\gamma$. 
At a fixed value of $\gamma$, the single and multi-pair production probability for aKTP and ppKTP are the same, independent of the pump power, as the different pumping powers act to equate the probabilities of generating $n$ photon pairs.
A hard limit on available pump power for multiple bulk PDC sources could restrict one's ability to maximise brightness.
Future scalable architectures however are likely to be based on waveguides, which typically require only $\mu$W of pumping power.
Gaussian PMFs can also be achieved in waveguide sources either through domain engineering, or via inter-modal Spontaneous Four Wave Mixing (SFWM) in a multi-mode waveguide~\cite{paesani2020near}.

Future improvements will target higher interference visibilities by modifying the PEF.
In this work, the PEF is a \(1.3~\textrm{ps}\) long sech$^2$ pulse, imposing a theoretical limit on the maximum visibility of 98.7\%.
However, it is possible to achieve up to 99.9\% visibility directly with our crystals by engineering the PEF~\cite{graffitti2018design}.
Modification of the PEF can be achieved using pump-shaping techniques~\cite{Chen:19}.
Additionally, further improvements may be obtained by exploring the interplay of spatial and spectral modes generated in a non-linearity engineered crystal~\cite{Bruno:14,Guerreiro:13}.

\bibliography{bib}

\newpage
\clearpage
\newpage

%%%%%%%%%%%%%%%%%%%%%
%%% supplementary %%% 
%%%%%%%%%%%%%%%%%%%%%

\renewcommand{\theequation}{S\arabic{equation}}
\renewcommand{\thefigure}{S\arabic{figure}}
\renewcommand{\thetable}{S\arabic{table}}
\renewcommand{\thesection}{S\Roman{section}}
\setcounter{equation}{0}
\setcounter{figure}{0}
\setcounter{table}{0}
\setcounter{section}{0}

\appendix
\onecolumngrid

\section*{\normalfont\large{Supplementary Material}}

\begin{figure}[h]
    \vspace{-1em}
    \centering\includegraphics[width =.65\columnwidth]{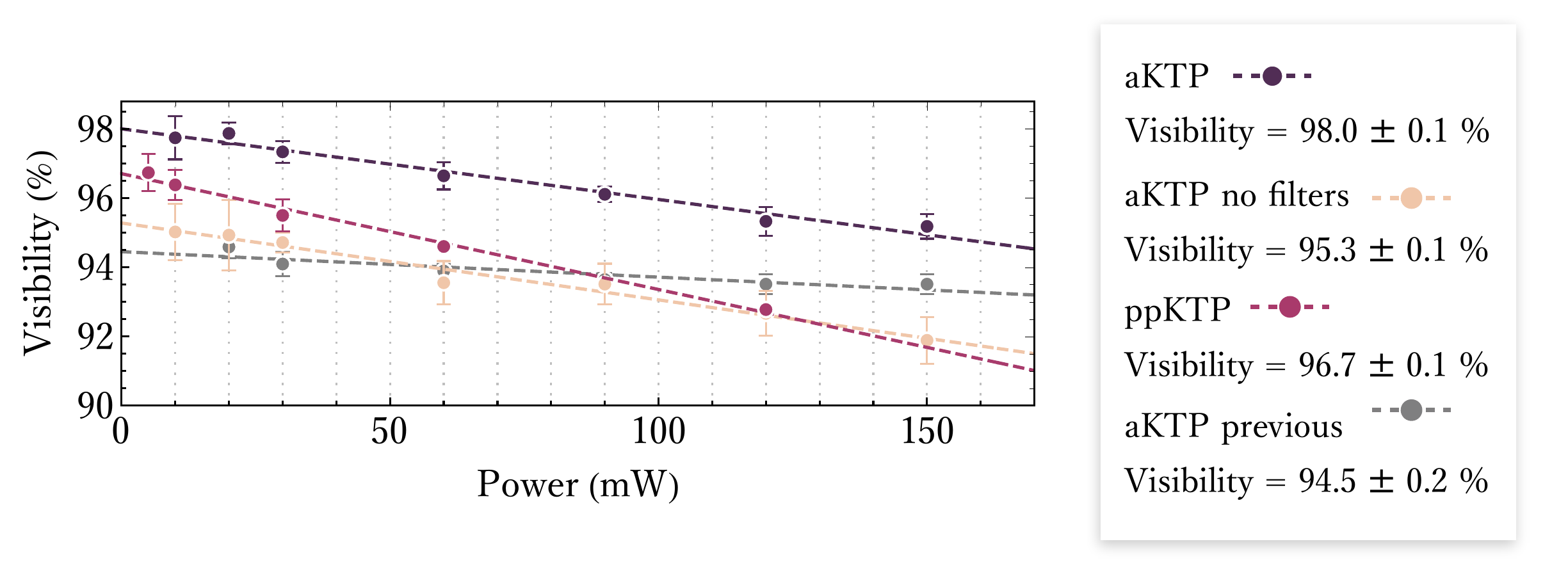}
    \vspace{-1em}
    \caption{Visibility power dependence. At each pump power setting, we measured five two-photon interferograms to obtain the average visibility shown here as solid data points. Error bars are taken as one standard deviation from these measurements. We then infer a minimum for spectral purity for a combination of crystals under certain filtering condition by fitting a linear function (dashed lines) to extract visibility under 0 pump power.}
    \label{fig:appendix_linearFits}
\end{figure}

For our crystal analysis we ran a visibility power dependence to extract linear fits allowing us to obtain a lower bound on the purity of the crystal and filter combination investigated. The full results, including measurements with our previous generation aKTP crystals are shown in in Figure~\ref{fig:appendix_linearFits}. We used the same optical setup for each analysis.

We also determined the interference visibilities for signal-idler interference. As mentioned in the main text, being able to interchangeably interfere daughter photons from PDC offers some additional capabilities when it comes to building multi-photon states. Under the same conditions as the results obtained with idler-idler interference we obtain a lower-bound purity of $97.0\pm0.1\%$, a $1\%$ decrease. 

\begin{figure*}[h]
    \centering\includegraphics[width=.7\columnwidth]{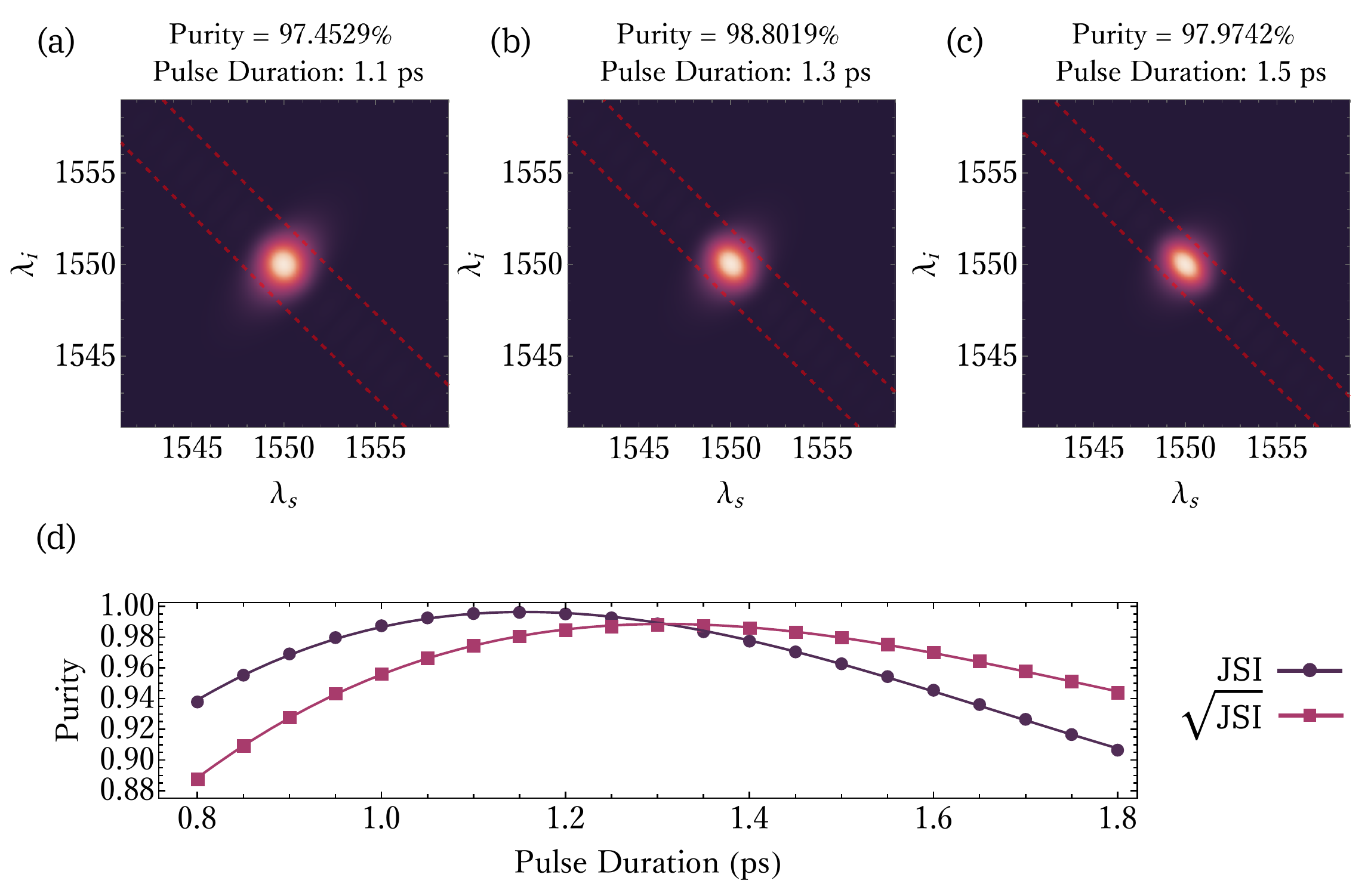}
    \vspace{-.5em}
    \caption{Theoretical simulations of photon purity as a function of varying pulse duration. A non-ideal pulse duration affects photon purity as the bandwidths of the PEF and PMF are not matched for pulse durations not 1.3~ps. (a), (b) and (c) depict the $\sqrt{\text{JSI}}$ from a range of pulse durations. Shorter pulse durations contribute towards spectral correlations along the diagonal, something visible in our reconstructed $\sqrt{\text{JSI}}$s. The red dash lines represent the width of the PEF corresponding to the pulse duration under investigation. (d) The effects of non-ideal pulse durations on photon purity. Analysing the range of purities derived from $\sqrt{\text{JSI}}$ and $\text{JSI}$ as a function of pump pulse duration.}
    \label{fig:appendix_alteringPEF}
\end{figure*}

Our efforts to consider why we witnessed lower purities in our experimental JSI analysis led to simulations into how pulse duration affects photon purity, the results of which are shown in Figure~\ref{fig:appendix_alteringPEF}. Maximum purities are achieved when the width of the PEF and PMF are matched. From the JSI reconstruction results, the elongation along the diagonal could have been caused by instability of our pulsed laser source, a reasonable argument as scans were run for hours at a time. Any drifting in pulse duration from ideal leads to a reduction in purity.

\end{document}

%% file: formatting.tex
\usepackage[utf8]{inputenc}
\usepackage[T1]{fontenc}
\usepackage{braket}
\usepackage{graphicx}
\usepackage{booktabs}
\usepackage{mathtools}	
\usepackage{amssymb}
\usepackage{amsmath}		
\usepackage{amsfonts}
\usepackage{mathrsfs}
\usepackage{color}
\usepackage{upgreek}
\usepackage{amsthm}
\usepackage{psfrag}
\usepackage{ifsym}
\usepackage{dsfont}
\usepackage{enumerate}
\usepackage{enumitem}
\usepackage{multirow} 
\usepackage{lipsum}
\usepackage{float}
\usepackage[separate-uncertainty=true]{siunitx}
\usepackage{setspace}
\usepackage[dvipsnames]{xcolor}
\usepackage[normalem]{ulem}

\usepackage{silence}
\usepackage{multirow}
\usepackage{booktabs}

\usepackage{hyperref}
\definecolor{refColour}{RGB}{0, 0, 0}
\definecolor{citeColour}{RGB}{0, 0, 0}
\definecolor{extraColour}{RGB}{0, 0, 0}
\hypersetup{colorlinks=true,linkcolor=refColour,citecolor=citeColour}

\usepackage{titlesec}
\titleformat*{\subsection}{\normalfont\centering}
\titleformat*{\section}{\normalfont\centering\large}

\newcommand{\one}{\leavevmode\hbox{\small1\normalsize\kern-.33em1}}

%% file: main.bbl
\begin{thebibliography}{52}
\expandafter\ifx\csname natexlab\endcsname\relax\def\natexlab#1{#1}\fi
\expandafter\ifx\csname bibnamefont\endcsname\relax
  \def\bibnamefont#1{#1}\fi
\expandafter\ifx\csname bibfnamefont\endcsname\relax
  \def\bibfnamefont#1{#1}\fi
\expandafter\ifx\csname citenamefont\endcsname\relax
  \def\citenamefont#1{#1}\fi
\expandafter\ifx\csname url\endcsname\relax
  \def\url#1{\texttt{#1}}\fi
\expandafter\ifx\csname urlprefix\endcsname\relax\def\urlprefix{URL }\fi
\providecommand{\bibinfo}[2]{#2}
\providecommand{\eprint}[2][]{\url{#2}}

\bibitem[{\citenamefont{Christ et~al.}(2013)\citenamefont{Christ, Fedrizzi,
  Hübel, Jennewein, and Silberhorn}}]{CHRIST2013351}
\bibinfo{author}{\bibfnamefont{A.}~\bibnamefont{Christ}},
  \bibinfo{author}{\bibfnamefont{A.}~\bibnamefont{Fedrizzi}},
  \bibinfo{author}{\bibfnamefont{H.}~\bibnamefont{Hübel}},
  \bibinfo{author}{\bibfnamefont{T.}~\bibnamefont{Jennewein}},
  \bibnamefont{and}
  \bibinfo{author}{\bibfnamefont{C.}~\bibnamefont{Silberhorn}}, in
  \emph{\bibinfo{booktitle}{Single-Photon Generation and Detection}}, edited by
  \bibinfo{editor}{\bibfnamefont{A.}~\bibnamefont{Migdall}},
  \bibinfo{editor}{\bibfnamefont{S.~V.} \bibnamefont{Polyakov}},
  \bibinfo{editor}{\bibfnamefont{J.}~\bibnamefont{Fan}}, \bibnamefont{and}
  \bibinfo{editor}{\bibfnamefont{J.~C.} \bibnamefont{Bienfang}}
  (\bibinfo{publisher}{Academic Press}, \bibinfo{year}{2013}),
  vol.~\bibinfo{volume}{45} of \emph{\bibinfo{series}{Experimental Methods in
  the Physical Sciences}}, pp. \bibinfo{pages}{351 -- 410},
  \urlprefix\url{http://www.sciencedirect.com/science/article/pii/B9780123876959000111}.

\bibitem[{\citenamefont{Slussarenko and Pryde}(2019)}]{slussarenko2019photonic}
\bibinfo{author}{\bibfnamefont{S.}~\bibnamefont{Slussarenko}} \bibnamefont{and}
  \bibinfo{author}{\bibfnamefont{G.~J.} \bibnamefont{Pryde}},
  \bibinfo{journal}{Applied Physics Reviews} \textbf{\bibinfo{volume}{6}},
  \bibinfo{pages}{041303} (\bibinfo{year}{2019}).

\bibitem[{\citenamefont{Pittman et~al.}(2002)\citenamefont{Pittman, Jacobs, and
  Franson}}]{PhysRevA.66.042303}
\bibinfo{author}{\bibfnamefont{T.~B.} \bibnamefont{Pittman}},
  \bibinfo{author}{\bibfnamefont{B.~C.} \bibnamefont{Jacobs}},
  \bibnamefont{and} \bibinfo{author}{\bibfnamefont{J.~D.}
  \bibnamefont{Franson}}, \bibinfo{journal}{Phys. Rev. A}
  \textbf{\bibinfo{volume}{66}}, \bibinfo{pages}{042303}
  (\bibinfo{year}{2002}),
  \urlprefix\url{https://link.aps.org/doi/10.1103/PhysRevA.66.042303}.

\bibitem[{\citenamefont{Migdall et~al.}(2002)\citenamefont{Migdall, Branning,
  and Castelletto}}]{PhysRevA.66.053805}
\bibinfo{author}{\bibfnamefont{A.~L.} \bibnamefont{Migdall}},
  \bibinfo{author}{\bibfnamefont{D.}~\bibnamefont{Branning}}, \bibnamefont{and}
  \bibinfo{author}{\bibfnamefont{S.}~\bibnamefont{Castelletto}},
  \bibinfo{journal}{Phys. Rev. A} \textbf{\bibinfo{volume}{66}},
  \bibinfo{pages}{053805} (\bibinfo{year}{2002}),
  \urlprefix\url{https://link.aps.org/doi/10.1103/PhysRevA.66.053805}.

\bibitem[{\citenamefont{Kaneda and Kwiat}(2019)}]{Kanedaeaaw8586}
\bibinfo{author}{\bibfnamefont{F.}~\bibnamefont{Kaneda}} \bibnamefont{and}
  \bibinfo{author}{\bibfnamefont{P.~G.} \bibnamefont{Kwiat}},
  \bibinfo{journal}{Science Advances} \textbf{\bibinfo{volume}{5}}
  (\bibinfo{year}{2019}),
  \urlprefix\url{https://advances.sciencemag.org/content/5/10/eaaw8586}.

\bibitem[{\citenamefont{Collins et~al.}(2013)\citenamefont{Collins, Xiong, Rey,
  Vo, He, Shahnia, Reardon, Krauss, Steel, Clark et~al.}}]{Collins2013}
\bibinfo{author}{\bibfnamefont{M.~J.} \bibnamefont{Collins}},
  \bibinfo{author}{\bibfnamefont{C.}~\bibnamefont{Xiong}},
  \bibinfo{author}{\bibfnamefont{I.~H.} \bibnamefont{Rey}},
  \bibinfo{author}{\bibfnamefont{T.~D.} \bibnamefont{Vo}},
  \bibinfo{author}{\bibfnamefont{J.}~\bibnamefont{He}},
  \bibinfo{author}{\bibfnamefont{S.}~\bibnamefont{Shahnia}},
  \bibinfo{author}{\bibfnamefont{C.}~\bibnamefont{Reardon}},
  \bibinfo{author}{\bibfnamefont{T.~F.} \bibnamefont{Krauss}},
  \bibinfo{author}{\bibfnamefont{M.~J.} \bibnamefont{Steel}},
  \bibinfo{author}{\bibfnamefont{A.~S.} \bibnamefont{Clark}},
  \bibnamefont{et~al.}, \bibinfo{journal}{Nature Communications}
  \textbf{\bibinfo{volume}{4}}, \bibinfo{pages}{2582} (\bibinfo{year}{2013}),
  ISSN \bibinfo{issn}{2041-1723},
  \urlprefix\url{https://doi.org/10.1038/ncomms3582}.

\bibitem[{\citenamefont{Kiyohara et~al.}(2016)\citenamefont{Kiyohara, Okamoto,
  and Takeuchi}}]{Kiyohara16muxWithPNR}
\bibinfo{author}{\bibfnamefont{T.}~\bibnamefont{Kiyohara}},
  \bibinfo{author}{\bibfnamefont{R.}~\bibnamefont{Okamoto}}, \bibnamefont{and}
  \bibinfo{author}{\bibfnamefont{S.}~\bibnamefont{Takeuchi}},
  \bibinfo{journal}{Opt. Express} \textbf{\bibinfo{volume}{24}},
  \bibinfo{pages}{27288} (\bibinfo{year}{2016}),
  \urlprefix\url{http://www.opticsexpress.org/abstract.cfm?URI=oe-24-24-27288}.

\bibitem[{\citenamefont{Francis-Jones et~al.}(2016)\citenamefont{Francis-Jones,
  Hoggarth, and Mosley}}]{Francis-Jones16muxFibre}
\bibinfo{author}{\bibfnamefont{R.~J.~A.} \bibnamefont{Francis-Jones}},
  \bibinfo{author}{\bibfnamefont{R.~A.} \bibnamefont{Hoggarth}},
  \bibnamefont{and} \bibinfo{author}{\bibfnamefont{P.~J.}
  \bibnamefont{Mosley}}, \bibinfo{journal}{Optica}
  \textbf{\bibinfo{volume}{3}}, \bibinfo{pages}{1270} (\bibinfo{year}{2016}),
  \urlprefix\url{http://www.osapublishing.org/optica/abstract.cfm?URI=optica-3-11-1270}.

\bibitem[{\citenamefont{Ma et~al.}(2011)\citenamefont{Ma, Zotter, Kofler,
  Jennewein, and Zeilinger}}]{Ma2011muxEOMs}
\bibinfo{author}{\bibfnamefont{X.-S.} \bibnamefont{Ma}},
  \bibinfo{author}{\bibfnamefont{S.}~\bibnamefont{Zotter}},
  \bibinfo{author}{\bibfnamefont{J.}~\bibnamefont{Kofler}},
  \bibinfo{author}{\bibfnamefont{T.}~\bibnamefont{Jennewein}},
  \bibnamefont{and}
  \bibinfo{author}{\bibfnamefont{A.}~\bibnamefont{Zeilinger}},
  \bibinfo{journal}{Physical Review A} \textbf{\bibinfo{volume}{83}},
  \bibinfo{pages}{043814} (\bibinfo{year}{2011}),
  \urlprefix\url{https://link.aps.org/doi/10.1103/PhysRevA.83.043814}.

\bibitem[{\citenamefont{Mendoza et~al.}(2016)\citenamefont{Mendoza, Santagati,
  Munns, Hemsley, Piekarek, Mart\'{i}n-L\'{o}pez, Marshall, Bonneau, Thompson,
  and O'Brien}}]{Mendoza16mux}
\bibinfo{author}{\bibfnamefont{G.~J.} \bibnamefont{Mendoza}},
  \bibinfo{author}{\bibfnamefont{R.}~\bibnamefont{Santagati}},
  \bibinfo{author}{\bibfnamefont{J.}~\bibnamefont{Munns}},
  \bibinfo{author}{\bibfnamefont{E.}~\bibnamefont{Hemsley}},
  \bibinfo{author}{\bibfnamefont{M.}~\bibnamefont{Piekarek}},
  \bibinfo{author}{\bibfnamefont{E.}~\bibnamefont{Mart\'{i}n-L\'{o}pez}},
  \bibinfo{author}{\bibfnamefont{G.~D.} \bibnamefont{Marshall}},
  \bibinfo{author}{\bibfnamefont{D.}~\bibnamefont{Bonneau}},
  \bibinfo{author}{\bibfnamefont{M.~G.} \bibnamefont{Thompson}},
  \bibnamefont{and} \bibinfo{author}{\bibfnamefont{J.~L.}
  \bibnamefont{O'Brien}}, \bibinfo{journal}{Optica}
  \textbf{\bibinfo{volume}{3}}, \bibinfo{pages}{127} (\bibinfo{year}{2016}),
  \urlprefix\url{http://www.osapublishing.org/optica/abstract.cfm?URI=optica-3-2-127}.

\bibitem[{\citenamefont{Broome et~al.}(2011)\citenamefont{Broome, Almeida,
  Fedrizzi, and White}}]{Broome:11}
\bibinfo{author}{\bibfnamefont{M.~A.} \bibnamefont{Broome}},
  \bibinfo{author}{\bibfnamefont{M.~P.} \bibnamefont{Almeida}},
  \bibinfo{author}{\bibfnamefont{A.}~\bibnamefont{Fedrizzi}}, \bibnamefont{and}
  \bibinfo{author}{\bibfnamefont{A.~G.} \bibnamefont{White}},
  \bibinfo{journal}{Opt. Express} \textbf{\bibinfo{volume}{19}},
  \bibinfo{pages}{22698} (\bibinfo{year}{2011}),
  \urlprefix\url{http://www.opticsexpress.org/abstract.cfm?URI=oe-19-23-22698}.

\bibitem[{\citenamefont{Meyer-Scott et~al.}(2020)\citenamefont{Meyer-Scott,
  Silberhorn, and Migdall}}]{doi:10.1063/5.0003320}
\bibinfo{author}{\bibfnamefont{E.}~\bibnamefont{Meyer-Scott}},
  \bibinfo{author}{\bibfnamefont{C.}~\bibnamefont{Silberhorn}},
  \bibnamefont{and} \bibinfo{author}{\bibfnamefont{A.}~\bibnamefont{Migdall}},
  \bibinfo{journal}{Review of Scientific Instruments}
  \textbf{\bibinfo{volume}{91}}, \bibinfo{pages}{041101}
  (\bibinfo{year}{2020}), \eprint{https://doi.org/10.1063/5.0003320},
  \urlprefix\url{https://doi.org/10.1063/5.0003320}.

\bibitem[{\citenamefont{Raussendorf et~al.}(2003)\citenamefont{Raussendorf,
  Browne, and Briegel}}]{PhysRevA.68.022312}
\bibinfo{author}{\bibfnamefont{R.}~\bibnamefont{Raussendorf}},
  \bibinfo{author}{\bibfnamefont{D.~E.} \bibnamefont{Browne}},
  \bibnamefont{and} \bibinfo{author}{\bibfnamefont{H.~J.}
  \bibnamefont{Briegel}}, \bibinfo{journal}{Phys. Rev. A}
  \textbf{\bibinfo{volume}{68}}, \bibinfo{pages}{022312}
  (\bibinfo{year}{2003}),
  \urlprefix\url{https://link.aps.org/doi/10.1103/PhysRevA.68.022312}.

\bibitem[{\citenamefont{Walther et~al.}(2005)\citenamefont{Walther, Resch,
  Rudolph, Schenck, Weinfurter, Vedral, Aspelmeyer, and Zeilinger}}]{oneWay}
\bibinfo{author}{\bibfnamefont{P.}~\bibnamefont{Walther}},
  \bibinfo{author}{\bibfnamefont{K.~J.} \bibnamefont{Resch}},
  \bibinfo{author}{\bibfnamefont{T.}~\bibnamefont{Rudolph}},
  \bibinfo{author}{\bibfnamefont{E.}~\bibnamefont{Schenck}},
  \bibinfo{author}{\bibfnamefont{H.}~\bibnamefont{Weinfurter}},
  \bibinfo{author}{\bibfnamefont{V.}~\bibnamefont{Vedral}},
  \bibinfo{author}{\bibfnamefont{M.}~\bibnamefont{Aspelmeyer}},
  \bibnamefont{and}
  \bibinfo{author}{\bibfnamefont{A.}~\bibnamefont{Zeilinger}},
  \bibinfo{journal}{Nature} \textbf{\bibinfo{volume}{434}},
  \bibinfo{pages}{169} (\bibinfo{year}{2005}).

\bibitem[{\citenamefont{Broome et~al.}(2013)\citenamefont{Broome, Fedrizzi,
  Rahimi-Keshari, Dove, Aaronson, Ralph, and White}}]{Broome794}
\bibinfo{author}{\bibfnamefont{M.~A.} \bibnamefont{Broome}},
  \bibinfo{author}{\bibfnamefont{A.}~\bibnamefont{Fedrizzi}},
  \bibinfo{author}{\bibfnamefont{S.}~\bibnamefont{Rahimi-Keshari}},
  \bibinfo{author}{\bibfnamefont{J.}~\bibnamefont{Dove}},
  \bibinfo{author}{\bibfnamefont{S.}~\bibnamefont{Aaronson}},
  \bibinfo{author}{\bibfnamefont{T.~C.} \bibnamefont{Ralph}}, \bibnamefont{and}
  \bibinfo{author}{\bibfnamefont{A.~G.} \bibnamefont{White}},
  \bibinfo{journal}{Science} \textbf{\bibinfo{volume}{339}},
  \bibinfo{pages}{794} (\bibinfo{year}{2013}), ISSN \bibinfo{issn}{0036-8075},
  \urlprefix\url{https://science.sciencemag.org/content/339/6121/794}.

\bibitem[{\citenamefont{van~der Meer et~al.}(2020)\citenamefont{van~der Meer,
  Renema, Brecht, Silberhorn, and Pinkse}}]{PhysRevA.101.063821}
\bibinfo{author}{\bibfnamefont{R.}~\bibnamefont{van~der Meer}},
  \bibinfo{author}{\bibfnamefont{J.~J.} \bibnamefont{Renema}},
  \bibinfo{author}{\bibfnamefont{B.}~\bibnamefont{Brecht}},
  \bibinfo{author}{\bibfnamefont{C.}~\bibnamefont{Silberhorn}},
  \bibnamefont{and} \bibinfo{author}{\bibfnamefont{P.~W.~H.}
  \bibnamefont{Pinkse}}, \bibinfo{journal}{Phys. Rev. A}
  \textbf{\bibinfo{volume}{101}}, \bibinfo{pages}{063821}
  (\bibinfo{year}{2020}),
  \urlprefix\url{https://link.aps.org/doi/10.1103/PhysRevA.101.063821}.

\bibitem[{\citenamefont{Azuma et~al.}(2015)\citenamefont{Azuma, Tamaki, and
  Lo}}]{allOptical}
\bibinfo{author}{\bibfnamefont{K.}~\bibnamefont{Azuma}},
  \bibinfo{author}{\bibfnamefont{K.}~\bibnamefont{Tamaki}}, \bibnamefont{and}
  \bibinfo{author}{\bibfnamefont{H.-K.} \bibnamefont{Lo}},
  \bibinfo{journal}{Nature Communications} \textbf{\bibinfo{volume}{6}},
  \bibinfo{pages}{6787} (\bibinfo{year}{2015}).

\bibitem[{\citenamefont{Grice et~al.}(2001)\citenamefont{Grice, U'Ren, and
  Walmsley}}]{PhysRevA.64.063815}
\bibinfo{author}{\bibfnamefont{W.~P.} \bibnamefont{Grice}},
  \bibinfo{author}{\bibfnamefont{A.~B.} \bibnamefont{U'Ren}}, \bibnamefont{and}
  \bibinfo{author}{\bibfnamefont{I.~A.} \bibnamefont{Walmsley}},
  \bibinfo{journal}{Phys. Rev. A} \textbf{\bibinfo{volume}{64}},
  \bibinfo{pages}{063815} (\bibinfo{year}{2001}),
  \urlprefix\url{https://link.aps.org/doi/10.1103/PhysRevA.64.063815}.

\bibitem[{\citenamefont{U'Ren et~al.}(2006)\citenamefont{U'Ren, Silberhorn,
  Erdmann, Banaszek, Grice, Walmsley, and Raymer}}]{u2006generation}
\bibinfo{author}{\bibfnamefont{A.~B.} \bibnamefont{U'Ren}},
  \bibinfo{author}{\bibfnamefont{C.}~\bibnamefont{Silberhorn}},
  \bibinfo{author}{\bibfnamefont{R.}~\bibnamefont{Erdmann}},
  \bibinfo{author}{\bibfnamefont{K.}~\bibnamefont{Banaszek}},
  \bibinfo{author}{\bibfnamefont{W.~P.} \bibnamefont{Grice}},
  \bibinfo{author}{\bibfnamefont{I.~A.} \bibnamefont{Walmsley}},
  \bibnamefont{and} \bibinfo{author}{\bibfnamefont{M.~G.}
  \bibnamefont{Raymer}}, \bibinfo{journal}{arXiv preprint quant-ph/0611019}
  (\bibinfo{year}{2006}).

\bibitem[{\citenamefont{Mosley et~al.}(2008)\citenamefont{Mosley, Lundeen,
  Smith, Wasylczyk, U'Ren, Silberhorn, and Walmsley}}]{PhysRevLett.100.133601}
\bibinfo{author}{\bibfnamefont{P.~J.} \bibnamefont{Mosley}},
  \bibinfo{author}{\bibfnamefont{J.~S.} \bibnamefont{Lundeen}},
  \bibinfo{author}{\bibfnamefont{B.~J.} \bibnamefont{Smith}},
  \bibinfo{author}{\bibfnamefont{P.}~\bibnamefont{Wasylczyk}},
  \bibinfo{author}{\bibfnamefont{A.~B.} \bibnamefont{U'Ren}},
  \bibinfo{author}{\bibfnamefont{C.}~\bibnamefont{Silberhorn}},
  \bibnamefont{and} \bibinfo{author}{\bibfnamefont{I.~A.}
  \bibnamefont{Walmsley}}, \bibinfo{journal}{Phys. Rev. Lett.}
  \textbf{\bibinfo{volume}{100}}, \bibinfo{pages}{133601}
  (\bibinfo{year}{2008}),
  \urlprefix\url{https://link.aps.org/doi/10.1103/PhysRevLett.100.133601}.

\bibitem[{\citenamefont{Jin et~al.}(2014)\citenamefont{Jin, Shimizu, Wakui,
  Fujiwara, Yamashita, Miki, Terai, Wang, and Sasaki}}]{Jin:14}
\bibinfo{author}{\bibfnamefont{R.-B.} \bibnamefont{Jin}},
  \bibinfo{author}{\bibfnamefont{R.}~\bibnamefont{Shimizu}},
  \bibinfo{author}{\bibfnamefont{K.}~\bibnamefont{Wakui}},
  \bibinfo{author}{\bibfnamefont{M.}~\bibnamefont{Fujiwara}},
  \bibinfo{author}{\bibfnamefont{T.}~\bibnamefont{Yamashita}},
  \bibinfo{author}{\bibfnamefont{S.}~\bibnamefont{Miki}},
  \bibinfo{author}{\bibfnamefont{H.}~\bibnamefont{Terai}},
  \bibinfo{author}{\bibfnamefont{Z.}~\bibnamefont{Wang}}, \bibnamefont{and}
  \bibinfo{author}{\bibfnamefont{M.}~\bibnamefont{Sasaki}},
  \bibinfo{journal}{Opt. Express} \textbf{\bibinfo{volume}{22}},
  \bibinfo{pages}{11498} (\bibinfo{year}{2014}),
  \urlprefix\url{http://www.opticsexpress.org/abstract.cfm?URI=oe-22-10-11498}.

\bibitem[{\citenamefont{Jin et~al.}(2013)\citenamefont{Jin, Shimizu, Wakui,
  Benichi, and Sasaki}}]{Jin:13}
\bibinfo{author}{\bibfnamefont{R.-B.} \bibnamefont{Jin}},
  \bibinfo{author}{\bibfnamefont{R.}~\bibnamefont{Shimizu}},
  \bibinfo{author}{\bibfnamefont{K.}~\bibnamefont{Wakui}},
  \bibinfo{author}{\bibfnamefont{H.}~\bibnamefont{Benichi}}, \bibnamefont{and}
  \bibinfo{author}{\bibfnamefont{M.}~\bibnamefont{Sasaki}},
  \bibinfo{journal}{Opt. Express} \textbf{\bibinfo{volume}{21}},
  \bibinfo{pages}{10659} (\bibinfo{year}{2013}),
  \urlprefix\url{http://www.opticsexpress.org/abstract.cfm?URI=oe-21-9-10659}.

\bibitem[{\citenamefont{Greganti et~al.}(2018)\citenamefont{Greganti,
  Schiansky, Calafell, Procopio, Rozema, and Walther}}]{Greganti:18}
\bibinfo{author}{\bibfnamefont{C.}~\bibnamefont{Greganti}},
  \bibinfo{author}{\bibfnamefont{P.}~\bibnamefont{Schiansky}},
  \bibinfo{author}{\bibfnamefont{I.~A.} \bibnamefont{Calafell}},
  \bibinfo{author}{\bibfnamefont{L.~M.} \bibnamefont{Procopio}},
  \bibinfo{author}{\bibfnamefont{L.~A.} \bibnamefont{Rozema}},
  \bibnamefont{and} \bibinfo{author}{\bibfnamefont{P.}~\bibnamefont{Walther}},
  \bibinfo{journal}{Opt. Express} \textbf{\bibinfo{volume}{26}},
  \bibinfo{pages}{3286} (\bibinfo{year}{2018}),
  \urlprefix\url{http://www.opticsexpress.org/abstract.cfm?URI=oe-26-3-3286}.

\bibitem[{\citenamefont{Wang et~al.}(2016)\citenamefont{Wang, Chen, Li, Huang,
  Liu, Chen, Luo, Su, Wu, Li et~al.}}]{PhysRevLett.117.210502}
\bibinfo{author}{\bibfnamefont{X.-L.} \bibnamefont{Wang}},
  \bibinfo{author}{\bibfnamefont{L.-K.} \bibnamefont{Chen}},
  \bibinfo{author}{\bibfnamefont{W.}~\bibnamefont{Li}},
  \bibinfo{author}{\bibfnamefont{H.-L.} \bibnamefont{Huang}},
  \bibinfo{author}{\bibfnamefont{C.}~\bibnamefont{Liu}},
  \bibinfo{author}{\bibfnamefont{C.}~\bibnamefont{Chen}},
  \bibinfo{author}{\bibfnamefont{Y.-H.} \bibnamefont{Luo}},
  \bibinfo{author}{\bibfnamefont{Z.-E.} \bibnamefont{Su}},
  \bibinfo{author}{\bibfnamefont{D.}~\bibnamefont{Wu}},
  \bibinfo{author}{\bibfnamefont{Z.-D.} \bibnamefont{Li}},
  \bibnamefont{et~al.}, \bibinfo{journal}{Phys. Rev. Lett.}
  \textbf{\bibinfo{volume}{117}}, \bibinfo{pages}{210502}
  (\bibinfo{year}{2016}),
  \urlprefix\url{https://link.aps.org/doi/10.1103/PhysRevLett.117.210502}.

\bibitem[{\citenamefont{Bra\'{n}czyk et~al.}(2011)\citenamefont{Bra\'{n}czyk,
  Fedrizzi, Stace, Ralph, and White}}]{Branczyk:11}
\bibinfo{author}{\bibfnamefont{A.~M.} \bibnamefont{Bra\'{n}czyk}},
  \bibinfo{author}{\bibfnamefont{A.}~\bibnamefont{Fedrizzi}},
  \bibinfo{author}{\bibfnamefont{T.~M.} \bibnamefont{Stace}},
  \bibinfo{author}{\bibfnamefont{T.~C.} \bibnamefont{Ralph}}, \bibnamefont{and}
  \bibinfo{author}{\bibfnamefont{A.~G.} \bibnamefont{White}},
  \bibinfo{journal}{Opt. Express} \textbf{\bibinfo{volume}{19}},
  \bibinfo{pages}{55} (\bibinfo{year}{2011}),
  \urlprefix\url{http://www.opticsexpress.org/abstract.cfm?URI=oe-19-1-55}.

\bibitem[{\citenamefont{Graffitti
  et~al.}(2018{\natexlab{a}})\citenamefont{Graffitti, Kelly-Massicotte,
  Fedrizzi, and Bra{\'n}czyk}}]{graffitti2018design}
\bibinfo{author}{\bibfnamefont{F.}~\bibnamefont{Graffitti}},
  \bibinfo{author}{\bibfnamefont{J.}~\bibnamefont{Kelly-Massicotte}},
  \bibinfo{author}{\bibfnamefont{A.}~\bibnamefont{Fedrizzi}}, \bibnamefont{and}
  \bibinfo{author}{\bibfnamefont{A.~M.} \bibnamefont{Bra{\'n}czyk}},
  \bibinfo{journal}{Physical Review A} \textbf{\bibinfo{volume}{98}},
  \bibinfo{pages}{053811} (\bibinfo{year}{2018}{\natexlab{a}}).

\bibitem[{\citenamefont{Bra{\'{n}}czyk
  et~al.}(2010)\citenamefont{Bra{\'{n}}czyk, Ralph, Helwig, and
  Silberhorn}}]{Bra_czyk_2010}
\bibinfo{author}{\bibfnamefont{A.~M.} \bibnamefont{Bra{\'{n}}czyk}},
  \bibinfo{author}{\bibfnamefont{T.~C.} \bibnamefont{Ralph}},
  \bibinfo{author}{\bibfnamefont{W.}~\bibnamefont{Helwig}}, \bibnamefont{and}
  \bibinfo{author}{\bibfnamefont{C.}~\bibnamefont{Silberhorn}},
  \bibinfo{journal}{New Journal of Physics} \textbf{\bibinfo{volume}{12}},
  \bibinfo{pages}{063001} (\bibinfo{year}{2010}),
  \urlprefix\url{https://doi.org/10.1088%2F1367-2630%2F12%2F6%2F063001}.

\bibitem[{\citenamefont{Meyer-Scott et~al.}(2017)\citenamefont{Meyer-Scott,
  Montaut, Tiedau, Sansoni, Herrmann, Bartley, and
  Silberhorn}}]{PhysRevA.95.061803}
\bibinfo{author}{\bibfnamefont{E.}~\bibnamefont{Meyer-Scott}},
  \bibinfo{author}{\bibfnamefont{N.}~\bibnamefont{Montaut}},
  \bibinfo{author}{\bibfnamefont{J.}~\bibnamefont{Tiedau}},
  \bibinfo{author}{\bibfnamefont{L.}~\bibnamefont{Sansoni}},
  \bibinfo{author}{\bibfnamefont{H.}~\bibnamefont{Herrmann}},
  \bibinfo{author}{\bibfnamefont{T.~J.} \bibnamefont{Bartley}},
  \bibnamefont{and}
  \bibinfo{author}{\bibfnamefont{C.}~\bibnamefont{Silberhorn}},
  \bibinfo{journal}{Phys. Rev. A} \textbf{\bibinfo{volume}{95}},
  \bibinfo{pages}{061803} (\bibinfo{year}{2017}),
  \urlprefix\url{https://link.aps.org/doi/10.1103/PhysRevA.95.061803}.

\bibitem[{\citenamefont{Dixon et~al.}(2013)\citenamefont{Dixon, Shapiro, and
  Wong}}]{BenDixon:13}
\bibinfo{author}{\bibfnamefont{P.~B.} \bibnamefont{Dixon}},
  \bibinfo{author}{\bibfnamefont{J.~H.} \bibnamefont{Shapiro}},
  \bibnamefont{and} \bibinfo{author}{\bibfnamefont{F.~N.~C.}
  \bibnamefont{Wong}}, \bibinfo{journal}{Opt. Express}
  \textbf{\bibinfo{volume}{21}}, \bibinfo{pages}{5879} (\bibinfo{year}{2013}),
  \urlprefix\url{http://www.opticsexpress.org/abstract.cfm?URI=oe-21-5-5879}.

\bibitem[{\citenamefont{Chen et~al.}(2017)\citenamefont{Chen, Bo, Niu, Xu,
  Zhang, Shapiro, and Wong}}]{Chen:17}
\bibinfo{author}{\bibfnamefont{C.}~\bibnamefont{Chen}},
  \bibinfo{author}{\bibfnamefont{C.}~\bibnamefont{Bo}},
  \bibinfo{author}{\bibfnamefont{M.~Y.} \bibnamefont{Niu}},
  \bibinfo{author}{\bibfnamefont{F.}~\bibnamefont{Xu}},
  \bibinfo{author}{\bibfnamefont{Z.}~\bibnamefont{Zhang}},
  \bibinfo{author}{\bibfnamefont{J.~H.} \bibnamefont{Shapiro}},
  \bibnamefont{and} \bibinfo{author}{\bibfnamefont{F.~N.~C.}
  \bibnamefont{Wong}}, \bibinfo{journal}{Opt. Express}
  \textbf{\bibinfo{volume}{25}}, \bibinfo{pages}{7300} (\bibinfo{year}{2017}),
  \urlprefix\url{http://www.opticsexpress.org/abstract.cfm?URI=oe-25-7-7300}.

\bibitem[{\citenamefont{Chen et~al.}(2019)\citenamefont{Chen, Heyes, Hong, Niu,
  Lita, Gerrits, Nam, Shapiro, and Wong}}]{Chen:19}
\bibinfo{author}{\bibfnamefont{C.}~\bibnamefont{Chen}},
  \bibinfo{author}{\bibfnamefont{J.~E.} \bibnamefont{Heyes}},
  \bibinfo{author}{\bibfnamefont{K.-H.} \bibnamefont{Hong}},
  \bibinfo{author}{\bibfnamefont{M.~Y.} \bibnamefont{Niu}},
  \bibinfo{author}{\bibfnamefont{A.~E.} \bibnamefont{Lita}},
  \bibinfo{author}{\bibfnamefont{T.}~\bibnamefont{Gerrits}},
  \bibinfo{author}{\bibfnamefont{S.~W.} \bibnamefont{Nam}},
  \bibinfo{author}{\bibfnamefont{J.~H.} \bibnamefont{Shapiro}},
  \bibnamefont{and} \bibinfo{author}{\bibfnamefont{F.~N.~C.}
  \bibnamefont{Wong}}, \bibinfo{journal}{Opt. Express}
  \textbf{\bibinfo{volume}{27}}, \bibinfo{pages}{11626} (\bibinfo{year}{2019}),
  \urlprefix\url{http://www.opticsexpress.org/abstract.cfm?URI=oe-27-8-11626}.

\bibitem[{\citenamefont{Cui et~al.}(2019)\citenamefont{Cui, Arian, Guha,
  Peyghambarian, Zhuang, and Zhang}}]{PhysRevApplied.12.034059}
\bibinfo{author}{\bibfnamefont{C.}~\bibnamefont{Cui}},
  \bibinfo{author}{\bibfnamefont{R.}~\bibnamefont{Arian}},
  \bibinfo{author}{\bibfnamefont{S.}~\bibnamefont{Guha}},
  \bibinfo{author}{\bibfnamefont{N.}~\bibnamefont{Peyghambarian}},
  \bibinfo{author}{\bibfnamefont{Q.}~\bibnamefont{Zhuang}}, \bibnamefont{and}
  \bibinfo{author}{\bibfnamefont{Z.}~\bibnamefont{Zhang}},
  \bibinfo{journal}{Phys. Rev. Applied} \textbf{\bibinfo{volume}{12}},
  \bibinfo{pages}{034059} (\bibinfo{year}{2019}),
  \urlprefix\url{https://link.aps.org/doi/10.1103/PhysRevApplied.12.034059}.

\bibitem[{\citenamefont{Dosseva et~al.}(2016)\citenamefont{Dosseva, Cincio, and
  Bra\ifmmode~\acute{n}\else \'{n}\fi{}czyk}}]{Dosseva}
\bibinfo{author}{\bibfnamefont{A.}~\bibnamefont{Dosseva}},
  \bibinfo{author}{\bibfnamefont{L.}~\bibnamefont{Cincio}}, \bibnamefont{and}
  \bibinfo{author}{\bibfnamefont{A.~M.} \bibnamefont{Bra\ifmmode~\acute{n}\else
  \'{n}\fi{}czyk}}, \bibinfo{journal}{Phys. Rev. A}
  \textbf{\bibinfo{volume}{93}}, \bibinfo{pages}{013801}
  (\bibinfo{year}{2016}),
  \urlprefix\url{https://link.aps.org/doi/10.1103/PhysRevA.93.013801}.

\bibitem[{\citenamefont{Graffitti
  et~al.}(2018{\natexlab{b}})\citenamefont{Graffitti, Barrow, Proietti, Kundys,
  and Fedrizzi}}]{graffitti2018independent}
\bibinfo{author}{\bibfnamefont{F.}~\bibnamefont{Graffitti}},
  \bibinfo{author}{\bibfnamefont{P.}~\bibnamefont{Barrow}},
  \bibinfo{author}{\bibfnamefont{M.}~\bibnamefont{Proietti}},
  \bibinfo{author}{\bibfnamefont{D.}~\bibnamefont{Kundys}}, \bibnamefont{and}
  \bibinfo{author}{\bibfnamefont{A.}~\bibnamefont{Fedrizzi}},
  \bibinfo{journal}{Optica} \textbf{\bibinfo{volume}{5}}, \bibinfo{pages}{514}
  (\bibinfo{year}{2018}{\natexlab{b}}).

\bibitem[{\citenamefont{Tambasco et~al.}(2016)\citenamefont{Tambasco, Boes,
  Helt, Steel, and Mitchell}}]{Tambasco:16}
\bibinfo{author}{\bibfnamefont{J.-L.} \bibnamefont{Tambasco}},
  \bibinfo{author}{\bibfnamefont{A.}~\bibnamefont{Boes}},
  \bibinfo{author}{\bibfnamefont{L.~G.} \bibnamefont{Helt}},
  \bibinfo{author}{\bibfnamefont{M.~J.} \bibnamefont{Steel}}, \bibnamefont{and}
  \bibinfo{author}{\bibfnamefont{A.}~\bibnamefont{Mitchell}},
  \bibinfo{journal}{Opt. Express} \textbf{\bibinfo{volume}{24}},
  \bibinfo{pages}{19616} (\bibinfo{year}{2016}),
  \urlprefix\url{http://www.opticsexpress.org/abstract.cfm?URI=oe-24-17-19616}.

\bibitem[{\citenamefont{Graffitti et~al.}(2017)\citenamefont{Graffitti, Kundys,
  Reid, Bra{\'n}czyk, and Fedrizzi}}]{graffitti2017pure}
\bibinfo{author}{\bibfnamefont{F.}~\bibnamefont{Graffitti}},
  \bibinfo{author}{\bibfnamefont{D.}~\bibnamefont{Kundys}},
  \bibinfo{author}{\bibfnamefont{D.~T.} \bibnamefont{Reid}},
  \bibinfo{author}{\bibfnamefont{A.~M.} \bibnamefont{Bra{\'n}czyk}},
  \bibnamefont{and} \bibinfo{author}{\bibfnamefont{A.}~\bibnamefont{Fedrizzi}},
  \bibinfo{journal}{Quantum Science and Technology}
  \textbf{\bibinfo{volume}{2}}, \bibinfo{pages}{035001} (\bibinfo{year}{2017}).

\bibitem[{\citenamefont{Graffitti
  et~al.}(2020{\natexlab{a}})\citenamefont{Graffitti, Barrow, Pickston,
  Bra\ifmmode~\acute{n}\else \'{n}\fi{}czyk, and
  Fedrizzi}}]{PhysRevLett.124.053603}
\bibinfo{author}{\bibfnamefont{F.}~\bibnamefont{Graffitti}},
  \bibinfo{author}{\bibfnamefont{P.}~\bibnamefont{Barrow}},
  \bibinfo{author}{\bibfnamefont{A.}~\bibnamefont{Pickston}},
  \bibinfo{author}{\bibfnamefont{A.~M.} \bibnamefont{Bra\ifmmode~\acute{n}\else
  \'{n}\fi{}czyk}}, \bibnamefont{and}
  \bibinfo{author}{\bibfnamefont{A.}~\bibnamefont{Fedrizzi}},
  \bibinfo{journal}{Phys. Rev. Lett.} \textbf{\bibinfo{volume}{124}},
  \bibinfo{pages}{053603} (\bibinfo{year}{2020}{\natexlab{a}}),
  \urlprefix\url{https://link.aps.org/doi/10.1103/PhysRevLett.124.053603}.

\bibitem[{\citenamefont{Graffitti
  et~al.}(2020{\natexlab{b}})\citenamefont{Graffitti, D'Ambrosio, Proietti, Ho,
  Piccirillo, de~Lisio, Marrucci, and
  Fedrizzi}}]{graffitti2020hyperentanglement}
\bibinfo{author}{\bibfnamefont{F.}~\bibnamefont{Graffitti}},
  \bibinfo{author}{\bibfnamefont{V.}~\bibnamefont{D'Ambrosio}},
  \bibinfo{author}{\bibfnamefont{M.}~\bibnamefont{Proietti}},
  \bibinfo{author}{\bibfnamefont{J.}~\bibnamefont{Ho}},
  \bibinfo{author}{\bibfnamefont{B.}~\bibnamefont{Piccirillo}},
  \bibinfo{author}{\bibfnamefont{C.}~\bibnamefont{de~Lisio}},
  \bibinfo{author}{\bibfnamefont{L.}~\bibnamefont{Marrucci}}, \bibnamefont{and}
  \bibinfo{author}{\bibfnamefont{A.}~\bibnamefont{Fedrizzi}},
  \bibinfo{journal}{arXiv preprint arXiv:2006.01845}
  (\bibinfo{year}{2020}{\natexlab{b}}).

\bibitem[{\citenamefont{{Fejer} et~al.}(1992)\citenamefont{{Fejer}, {Magel},
  {Jundt}, and {Byer}}}]{161322}
\bibinfo{author}{\bibfnamefont{M.~M.} \bibnamefont{{Fejer}}},
  \bibinfo{author}{\bibfnamefont{G.~A.} \bibnamefont{{Magel}}},
  \bibinfo{author}{\bibfnamefont{D.~H.} \bibnamefont{{Jundt}}},
  \bibnamefont{and} \bibinfo{author}{\bibfnamefont{R.~L.}
  \bibnamefont{{Byer}}}, \bibinfo{journal}{IEEE Journal of Quantum Electronics}
  \textbf{\bibinfo{volume}{28}}, \bibinfo{pages}{2631} (\bibinfo{year}{1992}).

\bibitem[{\citenamefont{Quesada and Bra\ifmmode~\acute{n}\else
  \'{n}\fi{}czyk}(2018)}]{PhysRevA.98.043813}
\bibinfo{author}{\bibfnamefont{N.}~\bibnamefont{Quesada}} \bibnamefont{and}
  \bibinfo{author}{\bibfnamefont{A.~M.} \bibnamefont{Bra\ifmmode~\acute{n}\else
  \'{n}\fi{}czyk}}, \bibinfo{journal}{Phys. Rev. A}
  \textbf{\bibinfo{volume}{98}}, \bibinfo{pages}{043813}
  (\bibinfo{year}{2018}),
  \urlprefix\url{https://link.aps.org/doi/10.1103/PhysRevA.98.043813}.

\bibitem[{\citenamefont{Boyd}(2008)}]{boyd}
\bibinfo{author}{\bibfnamefont{R.~W.} \bibnamefont{Boyd}},
  \emph{\bibinfo{title}{Nonlinear Optics, Third Edition}}
  (\bibinfo{publisher}{Academic Press, Inc.}, \bibinfo{address}{USA},
  \bibinfo{year}{2008}), \bibinfo{edition}{3rd} ed., ISBN
  \bibinfo{isbn}{0123694701}.

\bibitem[{\citenamefont{Avenhaus et~al.}(2009)\citenamefont{Avenhaus, Eckstein,
  Mosley, and Silberhorn}}]{Avenhaus:09}
\bibinfo{author}{\bibfnamefont{M.}~\bibnamefont{Avenhaus}},
  \bibinfo{author}{\bibfnamefont{A.}~\bibnamefont{Eckstein}},
  \bibinfo{author}{\bibfnamefont{P.~J.} \bibnamefont{Mosley}},
  \bibnamefont{and}
  \bibinfo{author}{\bibfnamefont{C.}~\bibnamefont{Silberhorn}},
  \bibinfo{journal}{Opt. Lett.} \textbf{\bibinfo{volume}{34}},
  \bibinfo{pages}{2873} (\bibinfo{year}{2009}),
  \urlprefix\url{http://ol.osa.org/abstract.cfm?URI=ol-34-18-2873}.

\bibitem[{\citenamefont{Bennink}(2010)}]{BenninkRyanS2010OcGb}
\bibinfo{author}{\bibfnamefont{R.~S.} \bibnamefont{Bennink}},
  \bibinfo{journal}{Physical review. A, Atomic, molecular, and optical physics}
  \textbf{\bibinfo{volume}{81}} (\bibinfo{year}{2010}), ISSN
  \bibinfo{issn}{1094-1622}.

\bibitem[{\citenamefont{Grice et~al.}(2011)\citenamefont{Grice, Bennink,
  Goodman, and Ryan}}]{PhysRevA.83.023810}
\bibinfo{author}{\bibfnamefont{W.~P.} \bibnamefont{Grice}},
  \bibinfo{author}{\bibfnamefont{R.~S.} \bibnamefont{Bennink}},
  \bibinfo{author}{\bibfnamefont{D.~S.} \bibnamefont{Goodman}},
  \bibnamefont{and} \bibinfo{author}{\bibfnamefont{A.~T.} \bibnamefont{Ryan}},
  \bibinfo{journal}{Phys. Rev. A} \textbf{\bibinfo{volume}{83}},
  \bibinfo{pages}{023810} (\bibinfo{year}{2011}),
  \urlprefix\url{https://link.aps.org/doi/10.1103/PhysRevA.83.023810}.

\bibitem[{\citenamefont{Dixon et~al.}(2014)\citenamefont{Dixon, Rosenberg,
  Stelmakh, Grein, Bennink, Dauler, Kerman, Molnar, and
  Wong}}]{PhysRevA.90.043804}
\bibinfo{author}{\bibfnamefont{P.~B.} \bibnamefont{Dixon}},
  \bibinfo{author}{\bibfnamefont{D.}~\bibnamefont{Rosenberg}},
  \bibinfo{author}{\bibfnamefont{V.}~\bibnamefont{Stelmakh}},
  \bibinfo{author}{\bibfnamefont{M.~E.} \bibnamefont{Grein}},
  \bibinfo{author}{\bibfnamefont{R.~S.} \bibnamefont{Bennink}},
  \bibinfo{author}{\bibfnamefont{E.~A.} \bibnamefont{Dauler}},
  \bibinfo{author}{\bibfnamefont{A.~J.} \bibnamefont{Kerman}},
  \bibinfo{author}{\bibfnamefont{R.~J.} \bibnamefont{Molnar}},
  \bibnamefont{and} \bibinfo{author}{\bibfnamefont{F.~N.~C.}
  \bibnamefont{Wong}}, \bibinfo{journal}{Phys. Rev. A}
  \textbf{\bibinfo{volume}{90}}, \bibinfo{pages}{043804}
  (\bibinfo{year}{2014}),
  \urlprefix\url{https://link.aps.org/doi/10.1103/PhysRevA.90.043804}.

\bibitem[{\citenamefont{Jin et~al.}(2015)\citenamefont{Jin, Fujiwara,
  Yamashita, Miki, Terai, Wang, Wakui, Shimizu, and Sasaki}}]{JIN201547}
\bibinfo{author}{\bibfnamefont{R.-B.} \bibnamefont{Jin}},
  \bibinfo{author}{\bibfnamefont{M.}~\bibnamefont{Fujiwara}},
  \bibinfo{author}{\bibfnamefont{T.}~\bibnamefont{Yamashita}},
  \bibinfo{author}{\bibfnamefont{S.}~\bibnamefont{Miki}},
  \bibinfo{author}{\bibfnamefont{H.}~\bibnamefont{Terai}},
  \bibinfo{author}{\bibfnamefont{Z.}~\bibnamefont{Wang}},
  \bibinfo{author}{\bibfnamefont{K.}~\bibnamefont{Wakui}},
  \bibinfo{author}{\bibfnamefont{R.}~\bibnamefont{Shimizu}}, \bibnamefont{and}
  \bibinfo{author}{\bibfnamefont{M.}~\bibnamefont{Sasaki}},
  \bibinfo{journal}{Optics Communications} \textbf{\bibinfo{volume}{336}},
  \bibinfo{pages}{47 } (\bibinfo{year}{2015}), ISSN \bibinfo{issn}{0030-4018},
  \urlprefix\url{http://www.sciencedirect.com/science/article/pii/S0030401814008803}.

\bibitem[{\citenamefont{Jizan et~al.}(2016)\citenamefont{Jizan, Bell, Helt,
  Bedoya, Xiong, and Eggleton}}]{Jizan:16}
\bibinfo{author}{\bibfnamefont{I.}~\bibnamefont{Jizan}},
  \bibinfo{author}{\bibfnamefont{B.}~\bibnamefont{Bell}},
  \bibinfo{author}{\bibfnamefont{L.~G.} \bibnamefont{Helt}},
  \bibinfo{author}{\bibfnamefont{A.~C.} \bibnamefont{Bedoya}},
  \bibinfo{author}{\bibfnamefont{C.}~\bibnamefont{Xiong}}, \bibnamefont{and}
  \bibinfo{author}{\bibfnamefont{B.~J.} \bibnamefont{Eggleton}},
  \bibinfo{journal}{Opt. Lett.} \textbf{\bibinfo{volume}{41}},
  \bibinfo{pages}{4803} (\bibinfo{year}{2016}),
  \urlprefix\url{http://ol.osa.org/abstract.cfm?URI=ol-41-20-4803}.

\bibitem[{\citenamefont{Law et~al.}(2000)\citenamefont{Law, Walmsley, and
  Eberly}}]{PhysRevLett.84.5304}
\bibinfo{author}{\bibfnamefont{C.~K.} \bibnamefont{Law}},
  \bibinfo{author}{\bibfnamefont{I.~A.} \bibnamefont{Walmsley}},
  \bibnamefont{and} \bibinfo{author}{\bibfnamefont{J.~H.}
  \bibnamefont{Eberly}}, \bibinfo{journal}{Phys. Rev. Lett.}
  \textbf{\bibinfo{volume}{84}}, \bibinfo{pages}{5304} (\bibinfo{year}{2000}),
  \urlprefix\url{https://link.aps.org/doi/10.1103/PhysRevLett.84.5304}.

\bibitem[{\citenamefont{Zhong et~al.}(2020)\citenamefont{Zhong, Wang, Deng,
  Chen, Peng, Luo, Qin, Wu, Ding, Hu et~al.}}]{Zhong1460}
\bibinfo{author}{\bibfnamefont{H.-S.} \bibnamefont{Zhong}},
  \bibinfo{author}{\bibfnamefont{H.}~\bibnamefont{Wang}},
  \bibinfo{author}{\bibfnamefont{Y.-H.} \bibnamefont{Deng}},
  \bibinfo{author}{\bibfnamefont{M.-C.} \bibnamefont{Chen}},
  \bibinfo{author}{\bibfnamefont{L.-C.} \bibnamefont{Peng}},
  \bibinfo{author}{\bibfnamefont{Y.-H.} \bibnamefont{Luo}},
  \bibinfo{author}{\bibfnamefont{J.}~\bibnamefont{Qin}},
  \bibinfo{author}{\bibfnamefont{D.}~\bibnamefont{Wu}},
  \bibinfo{author}{\bibfnamefont{X.}~\bibnamefont{Ding}},
  \bibinfo{author}{\bibfnamefont{Y.}~\bibnamefont{Hu}}, \bibnamefont{et~al.},
  \bibinfo{journal}{Science} \textbf{\bibinfo{volume}{370}},
  \bibinfo{pages}{1460} (\bibinfo{year}{2020}), ISSN \bibinfo{issn}{0036-8075},
  \urlprefix\url{https://science.sciencemag.org/content/370/6523/1460}.

\bibitem[{\citenamefont{Paesani et~al.}(2020)\citenamefont{Paesani, Borghi,
  Signorini, Ma{\"\i}nos, Pavesi, and Laing}}]{paesani2020near}
\bibinfo{author}{\bibfnamefont{S.}~\bibnamefont{Paesani}},
  \bibinfo{author}{\bibfnamefont{M.}~\bibnamefont{Borghi}},
  \bibinfo{author}{\bibfnamefont{S.}~\bibnamefont{Signorini}},
  \bibinfo{author}{\bibfnamefont{A.}~\bibnamefont{Ma{\"\i}nos}},
  \bibinfo{author}{\bibfnamefont{L.}~\bibnamefont{Pavesi}}, \bibnamefont{and}
  \bibinfo{author}{\bibfnamefont{A.}~\bibnamefont{Laing}},
  \bibinfo{journal}{Nature communications} \textbf{\bibinfo{volume}{11}},
  \bibinfo{pages}{1} (\bibinfo{year}{2020}).

\bibitem[{\citenamefont{Bruno et~al.}(2014)\citenamefont{Bruno, Martin,
  Guerreiro, Sanguinetti, and Thew}}]{Bruno:14}
\bibinfo{author}{\bibfnamefont{N.}~\bibnamefont{Bruno}},
  \bibinfo{author}{\bibfnamefont{A.}~\bibnamefont{Martin}},
  \bibinfo{author}{\bibfnamefont{T.}~\bibnamefont{Guerreiro}},
  \bibinfo{author}{\bibfnamefont{B.}~\bibnamefont{Sanguinetti}},
  \bibnamefont{and} \bibinfo{author}{\bibfnamefont{R.~T.} \bibnamefont{Thew}},
  \bibinfo{journal}{Opt. Express} \textbf{\bibinfo{volume}{22}},
  \bibinfo{pages}{17246} (\bibinfo{year}{2014}),
  \urlprefix\url{http://www.opticsexpress.org/abstract.cfm?URI=oe-22-14-17246}.

\bibitem[{\citenamefont{Guerreiro et~al.}(2013)\citenamefont{Guerreiro, Martin,
  Sanguinetti, Bruno, Zbinden, and Thew}}]{Guerreiro:13}
\bibinfo{author}{\bibfnamefont{T.}~\bibnamefont{Guerreiro}},
  \bibinfo{author}{\bibfnamefont{A.}~\bibnamefont{Martin}},
  \bibinfo{author}{\bibfnamefont{B.}~\bibnamefont{Sanguinetti}},
  \bibinfo{author}{\bibfnamefont{N.}~\bibnamefont{Bruno}},
  \bibinfo{author}{\bibfnamefont{H.}~\bibnamefont{Zbinden}}, \bibnamefont{and}
  \bibinfo{author}{\bibfnamefont{R.~T.} \bibnamefont{Thew}},
  \bibinfo{journal}{Opt. Express} \textbf{\bibinfo{volume}{21}},
  \bibinfo{pages}{27641} (\bibinfo{year}{2013}),
  \urlprefix\url{http://www.opticsexpress.org/abstract.cfm?URI=oe-21-23-27641}.

\end{thebibliography}
